\documentclass[preprint]{aastex631}
\usepackage{url} 
\def\aj{{AJ}}
\def\apj{{ApJ}}
\def\apjs{{ApJS}}
\def\apjl{{ApJL}}

\def\aap{{A\&A}}
\def\pasp{{PASP}}

\def\mnras{{MNRAS}}
\def\nat{{Nature}}

\usepackage{multirow}
\usepackage{graphicx}
\usepackage{amssymb}
\usepackage{amsmath}
\usepackage{epstopdf}
\usepackage{natbib}
\usepackage{color}
\usepackage{hyperref}
\hypersetup{colorlinks=true, linkcolor=blue, citecolor=blue,urlcolor=blue}
\usepackage[all]{hypcap}
\usepackage[flushleft]{threeparttable}

\graphicspath{{./}{Figures/}}

\begin{document}

%
%


\title{The Nominal Range of Rocky Planet Masses, Radii, Surface Gravities and Bulk Densities}

\author[0000-0001-8991-3110]{C.T. Unterborn}
\affiliation{Space Sector, Southwest Research Institute}

\author[0000-0002-1571-0836]{S.J. Desch}
\affiliation{School of Earth and Space Exploration, Arizona State University}

\author[0000-0003-1231-2389]{J. Haldemann}
\affiliation{Department of Space Research \& Planetary Sciences, University of Bern, Switzerland}

\author{A. Lorenzo}
\affiliation{School of Earth and Space Exploration, Arizona State University} 

\author[0000-0003-3570-422X]{J. G. Schulze}
\affiliation{School of Earth Science, The Ohio State University}

\author[0000-0003-0595-5132]{N. R. Hinkel}
\affiliation{Space Sector, Southwest Research Institute}

\author[0000-0001-5753-2532]{W.R.Panero}
\affiliation{School of Earth Science, The Ohio State University}

\correspondingauthor{Cayman Unterborn}
\email{cayman.unterborn@swri.org}

\begin{abstract}
The two primary observable quantities of an exoplanet--its mass and radius--alone are not sufficient to probe a rocky exoplanet's interior composition and mineralogy. To overcome this, host-star abundances of the primary planet-building elements (Mg, Si, Fe) are typically used as a proxy for the planet's bulk composition. The majority of small exoplanet hosts, however, do not have available abundance data. Here we present the open-source ExoPlex mass-radius-composition solver. Unlike previous open-source mass-radius solvers, ExoPlex calculates the core chemistry and equilibrium mantle mineralogy for a bulk composition, including effects of mantle FeO content, core light elements and surface water/ice. We utilize ExoPlex to calculate the planetary radii, surface gravities and bulk densities for 10$^6$ model planets up to 2 R$_\oplus$ across these geochemistries, adopting the distribution of FGK stellar abundances to estimate of the range of bulk exoplanet compositions. We outline the 99.7\% distribution of radii, surface gravity and bulk densities that define planets as ``nominally rocky.'' Planets outside this range require compositions outside those expected from stellar abundance data, likely making them either Fe-enriched super-Mercuries, or volatile-enriched mini-Neptunes. We apply our classification scheme to a sample of 85 well-resolved exoplanets without available host-star abundances. We estimate only 9 planets are within the ``nominally rocky planet zone'' at $>70\%$ confidence, while $\sim$20\% and $\sim$30\% of this sample can be reasonably classified as super-Mercuries or volatile-rich, respectively. Our results provide observers with a self-consistent way to broadly classify a planet as likely rocky, Mercury-like or volatile-enriched, using mass and radius measurements alone.
\end{abstract}

\section{Introduction} 

The compositions of small ($R \leq 2 \, R_{\oplus}$) exoplanets provide considerable information about their formation \citep[e.g.,][]{Unter18, Adibekyan21}, interior dynamics \citep{Ballmer2017, Spaargaren2020}, and potential habitability \citep[e.g., ][]{Unter22}. 
Attempts to quantify the composition of small exoplanets has been ongoing for over a decade, by attempting to match a planet's measured density to those predicted by mass-radius models \citep[e.g., ][]{Valencia2006, Seag07, Valencia2007, Valencia2007b, Zeng08, Dorn15, Unter16, Unter18, Huang22}.
Recent work has used mass-radius modeling to identify high-density super-Mercuries \citep[e.g., ][]{Bonomo19} as well as small planets that contain significant surface water \citep[e.g.,][]{Unter14} or thick atmosphere \citep[e.g, ][]{Brink22}, which lower their bulk densities considerably, making them water-worlds and mini-Neptunes, respectively. 

For more intermediate-density exoplanets, both forward \citep{Unter16} and inverse \citep{Dorn15} mass-radius models show considerable degeneracy when inferring basic aspects of a planet's bulk interior composition, chemistry and structure via mass and radius measurements alone. 
To break this degeneracy, the abundances of the primary rocky planet-building elements Mg, Si, Al, Ca, and Fe from the host-star are often assumed to be a one-to-one proxy for that of the exoplanet \citep[e.g.,][]{Thia15, Dorn15, Unter16}. 
This assumption is a good match for the Earth and Sun \citep{Dorn15, Unter16} as volatility need not be considered for these elements \citep{Wang19}. 
Other planets have inferred compositions that are statistically indistinguishable from those of their host stars \citep{Schulze21}, while others have densities consistent with having significant Fe-enrichment \citep{Bonomo19}, or extensive volatile layers \citep{Brink22}. 
Unfortunately, host-star abundances are not available for the majority of individual exoplanets, limiting our ability to distinguish them as rocky, iron or gas-rich exoplanets. 
Abundances from FGK stars indicate a broad range of possible abundances of primary rocky-planet building elements relative to the Sun and Earth \citep{Brew16, Hink17,Unter19}. 
This wide range of stellar abundances can provide us with reasonable bounds on a planet's interior bulk composition despite the dearth of exoplanet host-star abundances. 
To first-order, changes in a planet's bulk composition affects the relative size of the planet's core to its mantle \citep[e.g.,][]{Unter19}, and its interior mineralogy \citep[e.g.,][]{Hink18}. 
Changes in the relative mass of the central Fe-core relative to the planet's mantle has the largest effect on a planet's bulk density \citep[e.g., ][]{Valencia2006, rogers_2002_aa, Dorn15,Unter16}.
Chemical processes too can affect a planet's bulk density. 
Examples include the relative fraction of iron that is removed from the core to create oxidized mantle FeO \citep[e.g., ][]{Schaefer2015} and the lowering of the core's bulk density due to the presence of alloyed light elements \citep[e.g., Si; ][]{Birch52, Schaefer2015, Unter16, SchlichtingYoung22}.
Thus while stellar abundances may inform the range of potential rocky exoplanet bulk densities due to compositional changes, 
chemical effects must also be taken into account.

Here we present ExoPlex\footnote{Available at \href{https://github.com/CaymanUnterborn/ExoPlex}{https://github.com/CaymanUnterborn/ExoPlex}}, the first open-source mass-radius calculator able to simultaneously consider mantle mineralogy and core chemistry across a wide range of planetary bulk compositions and oxidation states up to planets of $\sim 2 \, R_{\oplus}$. 
We begin by outlining the relevant equations, capabilities and structure of the ExoPlex model  and compare our results to previous mass-radius models (\S\ref{sec:methods}). 
We then utilize ExoPlex to quantify the likely range of rocky planet radii, surface gravities and bulk densities, as constrained by stellar abundance data (\S\ref{sec:stellar}) taking into account the effects of changes in core size (\S\ref{sec:baseline}), mantle FeO content (\S\ref{sec:FeO}), core light element budget (\S\ref{sec:core_LE}) and surface water contents (\S\ref{sec:water}).
We show that despite the considerable degeneracies in inferring the exact compositions of rocky exoplanets (\S\ref{sec:degen}), we are able to define a Nominally Rocky Planet Zone (NRPZ), that is the parameter space where planets are nominally rocky, versus those that require significant Fe- or volatile-enrichment, making them super-Mercuries and mini-Neptunes/water worlds, respectively (\S\ref{sec:nrpz}). 
We then demonstrate the efficacy of this broad classification scheme for a sample of small-exoplanet hosts by comparing their range of likely interior compositions based on mass and radius measurements alone to those predicted from their host stars' compositions (\S\ref{sec:final}).

\section{The ExoPlex Mass-Radius-Composition Calculator}
\label{sec:methods}
ExoPlex calculates a planet's mass or radius given its bulk composition of Mg, Si, Al, Ca and Fe (and O). It does this by simultaneously solving five coupled differential equations: the mass within a sphere,
\begin{equation}
    \frac{dm(r)}{dr} = 4\pi \,r^{2} \, \rho(r);
\end{equation}
the temperature-dependent Equation of State (EoS) for the constituent minerals,
\begin{equation}
\rho(r) = f(P(r),T(r));
\label{eq:eos}
\end{equation}
the equation of hydrostatic equilibrium,
\begin{equation}
\frac{dP(r)}{dr} = -\rho(r) \, g(r);
\label{eq:he}
\end{equation}
the adiabatic temperature profile,
\begin{equation}
\label{eq:temp}
\frac{dT(r)}{dr} = -\frac{\alpha(P, T) \, g(r)}{C_{P}(P, T)};
\end{equation}
and Gauss's law of gravity in one dimension,
\begin{equation}
\frac{1}{r^2} \, \frac{d}{dr} \, \left( r^{2} \, g(r) \right) = 4\pi G \, \rho(r).
\label{eq:gau}
\end{equation}
Here, $r$ is the radius; 
$m(r)$ is the mass;
$\rho(r)$ is the density; $P(r)$ and $T(r)$ are the pressure and temperature profiles; $\alpha(P, T)$ and $C_{P}(P,T)$ are the thermal expansivity and coefficient of specific heat at constant pressure for the minerals present at a location $r$ and pressure $P(r)$ and temperature $T(r)$; and $g(r)$ ($> 0$) is the acceleration due to gravity, where $G$ is the gravitational constant.
These are solved by utilizing \texttt{SciPy} finite difference methods \citep{scipy} on a grid of $N$ radial shells ($i = 0, 1, ... N$), with $r_0 = 0$ and $r_N$ being the outer radius of the planet. 
The mass enclosed within a radius $r_i$ is denoted $m_i$, and densities, compositions, and other quantities are defined at the same locations. 
The number of individual shells within the core, mantle or water layers are defined by the user.

Because the elemental composition of the bulk planet is an input, the relative mass of the core, mantle and water layers are also known through stoichiometry and mass balance (\S\ref{sec:bulk_comp}).
Therefore, each individual layer is assigned a mass equal to the total mass of the mantle, core, or water divided by the total number shells within that portion of the planet.
Planet radius is determined using the equation for the thickness of a spherical shell of given average density; for simplicity, we assume the average density of the layer is the arithmetic mean of the density at the bottom and top of the layer.
Given enough shells within the mantle and core, this assumption is robust, as density varies almost linearly within each layer.  
All models in this manuscript are run with 600, 500 and 700 core, mantle and water shells within each layer, respectively. These numbers of shells within each layer were chosen to better capture the locations of phase transitions in the mantle/water layers at depth. 
Models with as few as 300 shells in each layer also converge.
Starting with $r_0 = 0$, the radius of a layer, $r_i$ is found in terms of $r_{i-1}$ as:
\begin{equation}
    r_{i} = \left(\frac{3\left(m_{i}-m_{i-1}\right)}{4\pi\left(\rho_{i}+\rho_{i-1}\right)/2} + r_{i-1}^{3} \right)^{\frac{1}{3}}.
    \label{eq:rad}
\end{equation}
The value of the last shell yields the radius of the planet, $R = r_N$.

Given values of $\rho_i$, Equation~\ref{eq:gau} is easily integrated outward from $r=0$ (where $g = 0$) to find $g_i$ within each shell.
Then Equation~\ref{eq:he} is integrated inward from the surface, where $P(R) = P_N = 1$ bar, to find the pressure $P_i$ at each layer. 
To find the temperature $T_i$ at each layer requires integration of Equation~\ref{eq:temp} from the surface, where the temperature is defined to be $T(R) = T_{N} = T_{\rm pot}$, the potential temperature. 
ExoPlex does not model crustal material or an atmosphere, using the potential temperature instead.
As is usually defined, this is the temperature below the thermally conducting lithosphere, at the shallowest point where the temperature matches the adiabatic temperature profile of the mantle. 
In the case of a thick water layer, there are two potential temperatures that must be defined: one at the top of the water layer, and one at the top of the rocky mantle. 
No thermal boundary layer is assumed between the mantle and core.
This assumption is not true for the Earth as it is not in thermal equilibrium with the mantle \citep{Lay08}.
As an initial guess, ExoPlex requires only the mass, temperature and pressure profiles to be defined, as gravity and radius can be derived from these profiles.
ExoPlex assumes the mass within each layer is divided evenly within each layer based on the relative fraction of each based on the user's definition for the water layer and stoichiometry for the mantle and core (\S\ref{sec:bulk_comp}).
Initial pressure and temperature profiles assumed to be linear using the CMB temperature and pressure estimates from \citet{Unter19}.

For a given bulk composition and initial mass (or radius), Exoplex integrates Equations~\ref{eq:he}--\ref{eq:gau} to find $P(r)$ and $T(r)$; then updates the density $\rho(r)$ everywhere by applying Equation~\ref{eq:eos}.
The procedure is then repeated and iterated to convergence, defined to be when density changes between iterations $j-1$ and $j$ are $< 0.1\%$, i.e., 
\begin{equation}
    |1-\rho^{j}_{n}/\rho^{j-1}_{n}| \leq 10^{-3}
\end{equation}
in each shell $n$ of the modeled planet. 
For models where mass is input and a radius is output, ExoPlex simply compresses the planet by solving equations \ref{eq:eos}-\ref{eq:rad} until convergence is reached. 
Once convergence is reached, the equilibrium phase assemblage within each layer is determined by linearly interpolating within the 2-D phase diagram along the final pressure and temperature profile.
It is important to note that our determinations of $\rho(P,T), \alpha(P, T)$ and $C_P (P, T)$ already include the effects of various phases being present due to the bulk composition and oxidation state of the shell and this step simply quantifies the exact equilibrium phase proportions. 
This scheme allows us to capture the slow transition of various phases and polymorphs that occur across a range of depths.
This is particularly important as planet mass changes.
As planet mass increases, the pressure and temperature gradients within the mantle increases \citep{Unter19}, which causes the thickness of upper-mantle shells to thin (for planets with a similar number of shells).
Thus, while a planet's upper mantle may traverse the same pressure and temperature range with depth, the depth of the upper mantle will decrease, leading to less gradual phase transitions with depth. 
These phase transitions may be particularly important for the dynamics within the mantle due to density differences across the upper- and lower-most mantles \citep[e.g., ][]{Spaargaren2020}.

ExoPlex typically converges within 6--10 iterations, depending on planet size. 
For models where radius is given and mass is to be output, ExoPlex guesses a mass and compresses the planet until convergence is reached, then compares this radius to the desired one; the process is then repeated for different masses, and \texttt{SciPy} bracketing techniques \citep{scipy} are used to find a mass that matches the desired radius. 
ExoPlex converges in $\sim$0.33 and 0.42 seconds when averaged across 100 determinations on a single 3.2 GHz Apple M1 processor for a 1 and 10 M$_\oplus$ containing 300 mantle and core grids each, respectively. 
ExoPlex achieves this speed by combining pre-tabulated values of $\rho(P, T), \alpha(P,T)$, $C_{P}(P,T)$ and phase abundances across a wide range of pressures and temperatures for the mantle, core and water layers (\S\ref{sec:Mantle}--\S\ref{sec:water_eos}) with an optimized \texttt{qhull} triangulation \citep{qhull} and \texttt{Scipy} linear interpolation routine \citep{scipy}.

\subsection{Defining a planet's bulk composition}
\label{sec:bulk_comp}

Composition is defined within ExoPlex by inputting: the molar ratios of each element relative to Mg (e.g., Fe/Mg); the desired mantle mass fraction of Fe within the mantle as FeO; and inputting the desired mass fractions of Si, S and/or O within the Fe core.
For those elements only present in the mantle (Mg, Al, Ca) and the remaining Si and Fe not within the core, ExoPlex assumes all species are oxidized as MgO, Al$_2$O$_3$, CaO, SiO$_2$ and FeO. 
In this way, defining the abundances of Fe, Mg, Si, Ca, and Al, plus mantle FeO content and core S, Si and O content means we de facto define a bulk abundance of O by assuming the creation of oxides (e.g., SiO$_2$) and the incorporation of O into the core.
From these inputs, ExoPlex solves the 10 mass balance equations for each of the six elements, the amount of O needed to oxidize the mantle elements, and the amount of Si, S and O present in the core. 
This mass balance provides the relative mass fractions of each oxide in the mantle, the full core chemistry and the core mass fraction (CMF) for a given bulk composition.
When FeO is present in the mantle, ExoPlex conserves bulk Fe/Mg by placing some fraction of all Fe into the mantle, at the expense of the core's Fe budget. 
It also takes into account oxidation/reduction reactions when light elements enter the core at the expense of a mantle oxide (e.g., reaction \ref{eq:Si}). 
The details of this stoichiometry are described in \S\ref{sec:FeO} and \S\ref{sec:core_LE}. 
With the relative amounts of each elements present in the mantle and core set, ExoPlex then calculates the relative mass fractions of the mantle and core. 
The amount of water within a model is user-defined as the fraction of water relative to the planet's total mass. 

\subsection{Mantle mineralogy determination and equation of state}
\label{sec:Mantle}

ExoPlex treats the mantle bulk elemental composition as constant throughout the mantle. 
Within any shell $i$ defined as mantle, ExoPlex utilizes the \texttt{Perple\_X} thermodynamic equilibrium software \citep{Conn09} and adopting the underlying thermodynamic database of \citet{Stix11} to determine $\rho_i$, $C_{{\rm P},i}$ and $\alpha_i$ required in equations~\ref{eq:eos}--\ref{eq:rad} at the pressure and temperature within that shell, as well as the equilibrium mineralogy present in that shell.  
Users of ExoPlex have the option to directly call \texttt{Perple\_X} and calculate the mineralogies for the exact compositions and pressures and temperatures, if desired; but this method is computationally expensive.
Instead, ExoPlex typically relies on grids of mineral properties pre-computed using \texttt{Perple\_X}.
The results presented here rely on such grids of pre-computed values for different compositions, pressures, and temperatures, except where noted.
ExoPlex includes nearly 13,000 discrete grids of mineralogy and thermoelastic parameters that span molar compositions as follows: 
\begin{align*}
0.1 \leq \rm{Si/Mg} & \leq 2 \; \mathrm{\left(steps\: of\: 0.1\right)} \\ 
0.02 \leq \rm{Ca/Mg}& \leq 0.09 \; \mathrm{\left(steps\: of\: 0.01\right)}\\ 
0.04 \leq \rm{Al/Mg}& \leq 0.11 \; \mathrm{\left(steps\: of\: 0.01\right)}, 
\end{align*}
and individual mantle FeO contents of [0, 2, 4, 6, 8, 10, 15, 20] wt\%.
The ranges of molar ratios were chosen to span roughly the 2$\sigma$ ranges of each molar ratio represented in the stellar abundance data set \citep[e.g., ][]{Hink18}, while the range of FeO contents spans oxidation states from reduced bodies like Mercury \citep{Nitt18}, up to Mars-like planets \citep{Wanke1994}.
If a composition falls between any of these pre-computed input values, ExoPlex chooses the grid representing the nearest composition in each of the molar ratios. 
When wt\% FeO falls between the pre-computed input values, ExoPlex interpolates between the two nearest wt\% FeO grids using a molar weighting scheme for $\alpha(P,T)$ and individual phases, and a mass weighting scheme for $\rho(P,T)$ and $C_{P}(P,T)$.

For each of the combinations of composition (molar fractions and FeO content), there is a grid of values at pre-determined pressures and temperatures. 
These values include the mineralogy, density ($\rho$), specific heat at constant pressure ($C_{P}$), and thermal expansivity ($\alpha$).
A \texttt{qhull} triangulation \citep{qhull} and \texttt{Scipy} linear interpolation routine \citep{scipy} are used to convert these into the needed parameters at the exact pressure $P_i$ and temperature $T_i$ in equations~\ref{eq:he} and~\ref{eq:temp}, respectively.
For pressures where the majority of mantle phase transitions occur ($P \leq 140 \, \rm{GPa}$, $T \leq \, 3470 \, \rm{K}$), ExoPlex interpolates within a fine-grid of mineralogy and thermoelastic parameters calculated between 1 bar and 140 GPa, and between 1400 and 3500 K, with steps of 1 GPa and 90 K. 
For pressures and temperatures above this, a coarser grid is used for interpolation, due to the less complex mineralogy present at higher pressures and temperatures.
This high pressure/temperature grid spans 125 GPa to 2.8 TPa, and 1700 to 7000 K, with steps of $\sim$40 GPa and 225 K.

\begin{figure}
    \centering
    \includegraphics[width=\linewidth]{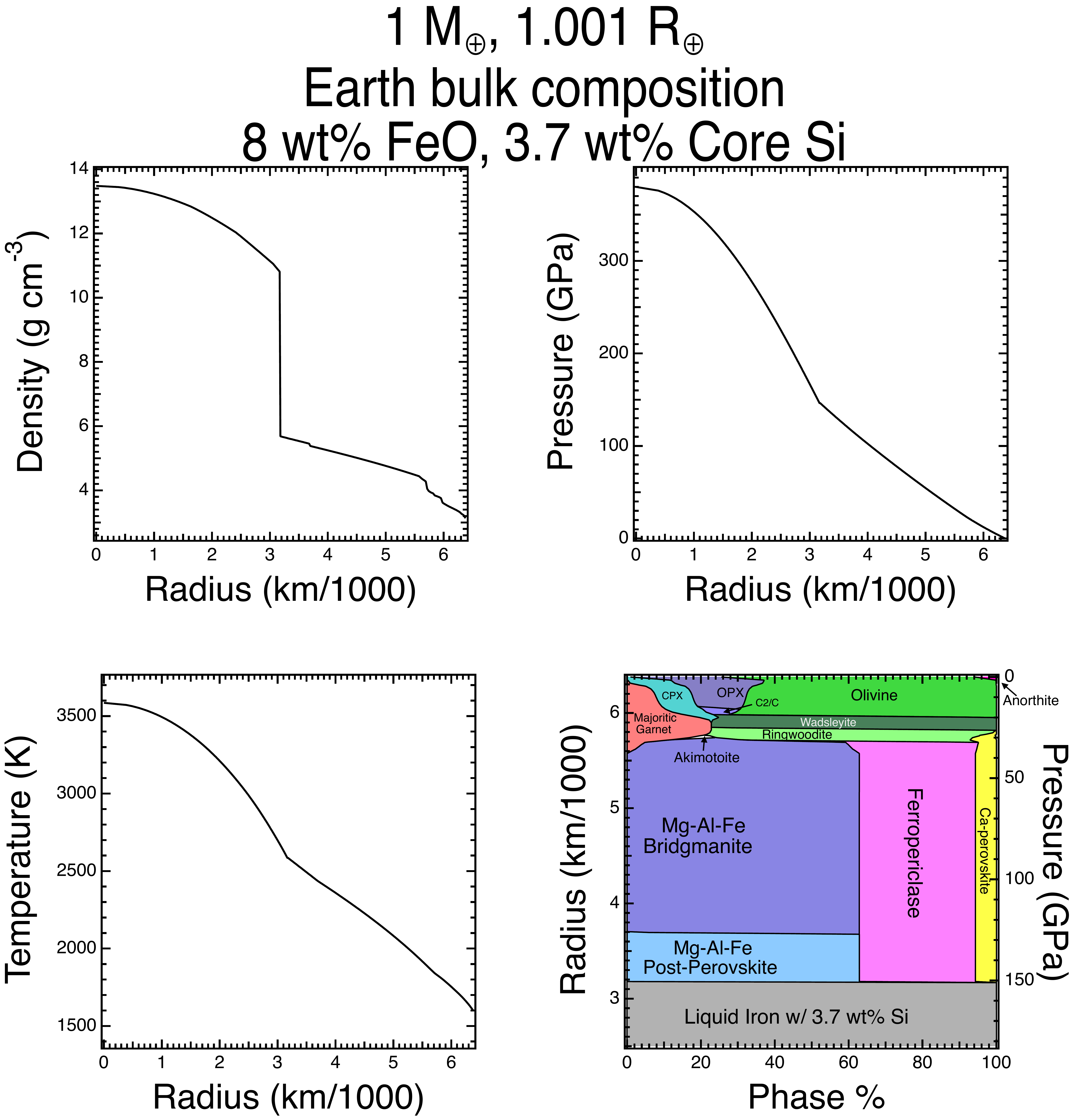}
    \caption{Density, pressure, adiabatic temperature and phase diagram for a 1 M$_\oplus$ planet for a planet of Earth composition \citep[Fe/Mg = 0.9, Si/Mg = 0.9, Ca/Mg = 0.07 and Al/Mg = 0.09;][] {McD03} and a mantle potential temperature of 1600 K derived using ExoPlex. This model assumes an Earth-like concentration of mantle FeO \citep[8 wt\%; ][]{McD03} and $\sim$3.7 wt\% Si within the core following reaction \ref{eq:Si}. This model yields a planet with a radius of 1.001 R$_\oplus$. OPX = orthopyroxene, CPX = clinopyroxene, C2/C = C2/C orthopyroxene.}
    \label{fig:phase_dia}
\end{figure}

\subsection{Core chemistry and equation of state}
\label{sec:core}

The mineralogy of the core is much simpler than that of the mantle. 
We adopt the equation of state of liquid iron of \citet{Ander94} as the default for ExoPlex to determine $\rho_i$, $C_{{\rm P},i}$ and $\alpha_i$ at each shell within the core.
This EoS is based on shock compression experiments up to 10 Mbar ($\sim 1$ TPa), which is roughly the central pressure of a $\sim \, 1.4 R_{\oplus}$ planet with an Earth-like core mass fraction (CMF) of 0.33 \citep{Unter19}. 
For pressures $> 1$ TPa, this EoS is extrapolated to the required pressure and temperature. 
For an individual run, ExoPlex interpolates within a $50 \times 50$ fine grid spanning 10 to 110 GPa and 1600 to 3600 K, applicable to smaller planets ($< 0.7 \, R_{\oplus}$), and a coarse $50 \times 50$ grid spanning 110 to 15,000 GPa and 1700 to 10,000 K, following the same routine as the mantle. 
These grids were generated using the BurnMan solver \citep{Cott14}.
As with the mantle EoS calculations, use of pre-computed grids drastically speeds up the calculation. 
We do not find any distinguishable difference in calculated density or adiabat profiles between cases where the EoS was calculated directly using BurnMan, versus interpolated using the grids. 
We explore the effects of adopting a solid Fe equation of state in \S\ref{sec:nrpz}.

ExoPlex supports the incorporation of likely light-alloying elements within the core (e.g., Si, O and S). 
These elements alloy with the Fe/Ni core due to the low- to moderate-pressure partitioning of elements between iron and silicates during core formation and differentiation in the magma ocean phase \citep[e.g., ][]{Schaefer2015} and/or during the formation of planetary materials in the disk \citep[e.g., ][]{Fisch20}.
While the identity of the light element budget in the Earth's core is debated \citep{Birch52, McD03}, these processes and the incorporation of light elements are likely to occur in rocky exoplanets as well \citep[e.g.,][]{Schaefer2015,SchlichtingYoung22}. 
Due to their lower molar mass than Fe, light elements can effectively lower the density of the core, which can lower the density of the planet entirely even if a planet's core mass fraction is conserved \citep{Birch52, Unter16, Unter18b,SchlichtingYoung22}.
Additionally, to first order, the incorporation of light elements (excluding H) does not significantly affect the compressibility of Fe, which could counteract the density decrease due to the lower molar weight of the core \citep{Poir94}.
For simplicity, then, ExoPlex simply reduces the interpolated density by the ratio of the calculated molar weight to that pure Fe depending on the molar fraction and species of light element present in to the core. 
This methodology is discussed further in \S\ref{sec:core_LE}.

\subsection{Water/Ice mineralogy determination and equation of state}
\label{sec:water_eos}
The mass fraction of the water layer is currently user-defined in terms of the fraction of the total planet mass.
ExoPlex adopts the open-source Seafreeze software package \citep{Journaux20} to determine both the phase and thermoelastic parameters of H$_2$O, assuming it is either liquid water, Ice Ih, II, III, V or VI. 
SeaFreeze utilizes a Gibbs Energy minimization framework to solve for $\rho$, $C_P$ and $\alpha$ for a given pressure and temperature, and is valid for pressures up to 2.3 GPa and temperatures between 220 and 500 K. 
For Ice VII we adopt the isothermal EoS of \citep{Journaux20} to determine $\rho (P,T)$, and the empirical equations of \citet{Asahara10} and \citet{Fei93} for $C_{P} (P,T)$ and $\alpha (P,T)$, respectively. 
As with the core and mantle, ExoPlex interpolates within a defined grid of pre-computed values, for pressures and temperatures below $\approx 3.7$ GPa and 500 K. 
For pressures and temperatures above this, we assume Ice VII is the dominant phase, and calculate the values of $\rho(P,T)$, $C_P(P,T)$ and $\alpha(P,T)$ directly. 
Our current models do not include Ice X, but will be included in future ExoPlex updates.

\subsection{Example ExoPlex output}
To demonstrate the capabilities of ExoPlex, we model a 1 M$_\oplus$ planet with an Earth bulk composition \citep{McD03} and Earth-like mantle FeO content \citep[8 wt\%; ][]{McD03}. 
Additionally, we assume a simultaneous oxidation/reduction reaction whereby the O needed to produce mantle FeO is taken from SiO$_2$ following reaction \ref{eq:Si}. 
This yields $\sim$3.7 wt\% of Si being present within the core, which reduces the core's density by $\sim$2\%.
This incorporation of Si into the core also lowers the mantle's molar Si/Mg from 0.9 to 0.83.
The resulting density, pressure, temperature and mineralogy profiles are shown in Figure \ref{fig:phase_dia}. 
For this composition and mass, ExoPlex produces a planet of $\sim1.001$ R$_\oplus$.

\subsection{Comparison to Previous Mass-Radius Models}
\begin{figure}
    \centering
    \includegraphics[width=\linewidth]{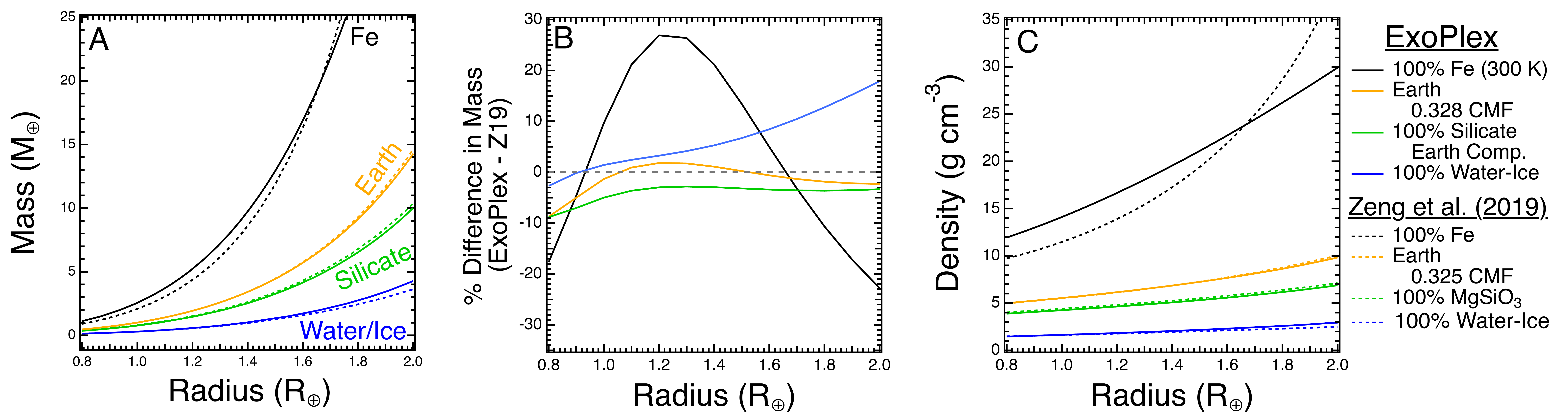}
    \caption{\textbf{A:} Modeled mass-radius curves for planets of pure Fe (black), silicate of Earth composition (green), water-ice (blue) and a 2-layer core/mantle planet with a roughly Earth-like core mass fraction (orange) using ExoPlex (solid) and those reported in \citet{Zeng19} (dashed). \textbf{B:} Percent difference in calculated mass between ExoPlex and \citet{Zeng19} for the same compositions. \textbf{C:} Planet density-radius curves for the same compositions.}
    \label{fig:MR}
\end{figure}

We use ExoPlex to calculate mass-radius and density-radius curves for various planets up to $2 \, R_{\oplus}$ (Figure~\ref{fig:MR}A): one made entirely of liquid Fe; one made entirely of silicate with Earth mantle-like composition \citep[Si/Mg = 0.9, Al/Mg = 0.09 and Ca/Mg = 0.07; ][]{McD03} without mantle FeO; one made of pure water-ice; and a two-layer planet with an Earth-like CMF ($\sim$0.33), with a liquid Fe core and Earth mantle composition. 
Comparing these models to those of \citet{Zeng19} (Z19), we find they agree to within 10\% up to 2 R$_\oplus$ for the pure-silicate and Earth models (Figure \ref{fig:MR}B).
ExoPlex's determination of pure-water/ice differs from Z19 at $>10\%$ above 1.7 R$_\oplus$, likely due to ExoPlex not including higher-order ice phases above ice-VII. 
The pure-Fe mass-radius curve, however, differs significantly between ExoPlex and Z19 (Figure~\ref{fig:MR}B).

A plot of bulk density vs. planet radius (Figure~\ref{fig:MR}C) shows that compared to ExoPlex, the Z19 models underestimate the density of a pure-Fe sphere for radii $< 1.6 \, R_{\oplus}$. 
For radii $> 1.6 \, R_{\oplus}$, the bulk density of Fe from Z19 begins to increase rapidly with planet radius (and central pressure), leading to predicted masses at $2 \, R_{\oplus}$ that are $\approx 12 M_{\oplus}$ (about 30\%) greater than those predicted by ExoPlex (Figure \ref{fig:MR}B). 
We attribute this difference to the fact that Z19 adopts a second-order Birch-Murnaghan (BM) EoS for liquid Fe \citep[taken from ][]{Zeng16} compared to the 4th order BM EoS adopted in ExoPlex. The 2nd order BM EoS neglects the reduction in Fe's compressibility (defined by the inverse of the bulk modulus) as pressure increases. 
By not adopting a higher-order EoS, the models of Z19 overestimate Fe's density at a given pressure, an effect that is exacerbated at higher planet masses with larger interior pressures. This will lead to the Z19 models likely underestimating a planet's CMF, which may be particularly important for quantifying the degree to which a super-Mercury is enriched in Fe.  

\section{Potential planet compositions as outlined by stellar abundance data}
\label{sec:stellar}
Small exoplanets span a wide range of densities. 
Mass-radius models tell us that these exoplanets are likely mixtures of rock, volatiles and iron of varying proportions \citep[e.g., ][]{Valencia2006, Seag07, Valencia2007, Valencia2007b, Zeng08, Dorn15, Unter16, Unter18, Huang22}.
Determining the exact proportions of rock, iron and/or volatiles is difficult, however, due to the considerable degeneracy of compositions that can match a planet's measured mass and radius. 
Often the assumption that the composition of the host-star roughly matches that of the planet is used to help break this degeneracy, at least for the elements that primarily make up rocky exoplanets (Mg, Si, Fe, Ca and Al). 
This assumption is well grounded as these elements, and the oxygen atoms required to create their constituent oxides (e.g., MgO and SiO$_2$), account for $\approx 92$\% of all atoms within the Earth, and $\approx 97\%$ of its mass \citep{McD03}. 
Furthermore, these elements all have equilibrium condensation temperatures greater than $\sim 1300$ K \citep{Lodd03} and are not expected to fractionate relative to each other during planetary formation processes \citep[e.g.,][]{Bond06, Bond10, Desch20}.
This means the relative ratios of these elements (e.g., Fe/Mg) should not differ significantly (more than about 10\%) between a host-star and its rocky exoplanet \citep[e.g.,][]{Bond06, Bond10, Thia15}. 
For comparison, the Earth, Mars and Sun match to within $\sim10\%$ in their relative ratios of these elements \citep{McD03, Lodd03, Wanke1994, Unter19}. 
Venus' bulk composition is not known. 

Abundance measurements for individual exoplanet host-stars are unfortunately sparse. 
The broader data set of abundances for FGK-type stars exhibits factor-of-two variations in the abundances of the refractory elements Fe, Mg and Si relative to solar \citep[e.g.,][]{Brew16, Hink18, Unter19}. 
These three elements, along with oxygen in their oxide forms (e.g., SiO$_2$), are the primary building blocks by mass for rocky exoplanets.
While other elements such as Ca, Al and Na are also likely to be present within rocky planets, they are an order-of-magnitude less abundant by mass than Mg, Fe and Si in the Sun and FGK-type stars \citep{Lodd03,Hink18}. 
While some enrichment of Ca, Al and Na relative to Fe, Mg and Si may be possible due to planetary formation processes \citep[e.g., ][]{Dorn19}, the bulk of a planet's observed mass is due to changes in the relative abundances of Fe, Mg and Si within the planet.
While some formation processes may change the relative abundances of Fe, Mg and Si between a host-star and its planets, the Earth, Sun and Mars show these differences can be small.  
Super-Mercury exoplanets, however, show a significant difference between a host-star's measured Fe abundance and the inferred value for the planet \citep[e.g., ][]{Bote18}. 
To first-order then, stellar abundances can provide us with a reasonable range of possible bulk planet compositions and can better allow us to identify planets that fall outside this range due to planetary formation processes. 

\begin{figure}[t]
    \centering
    \includegraphics[width=0.5\linewidth]{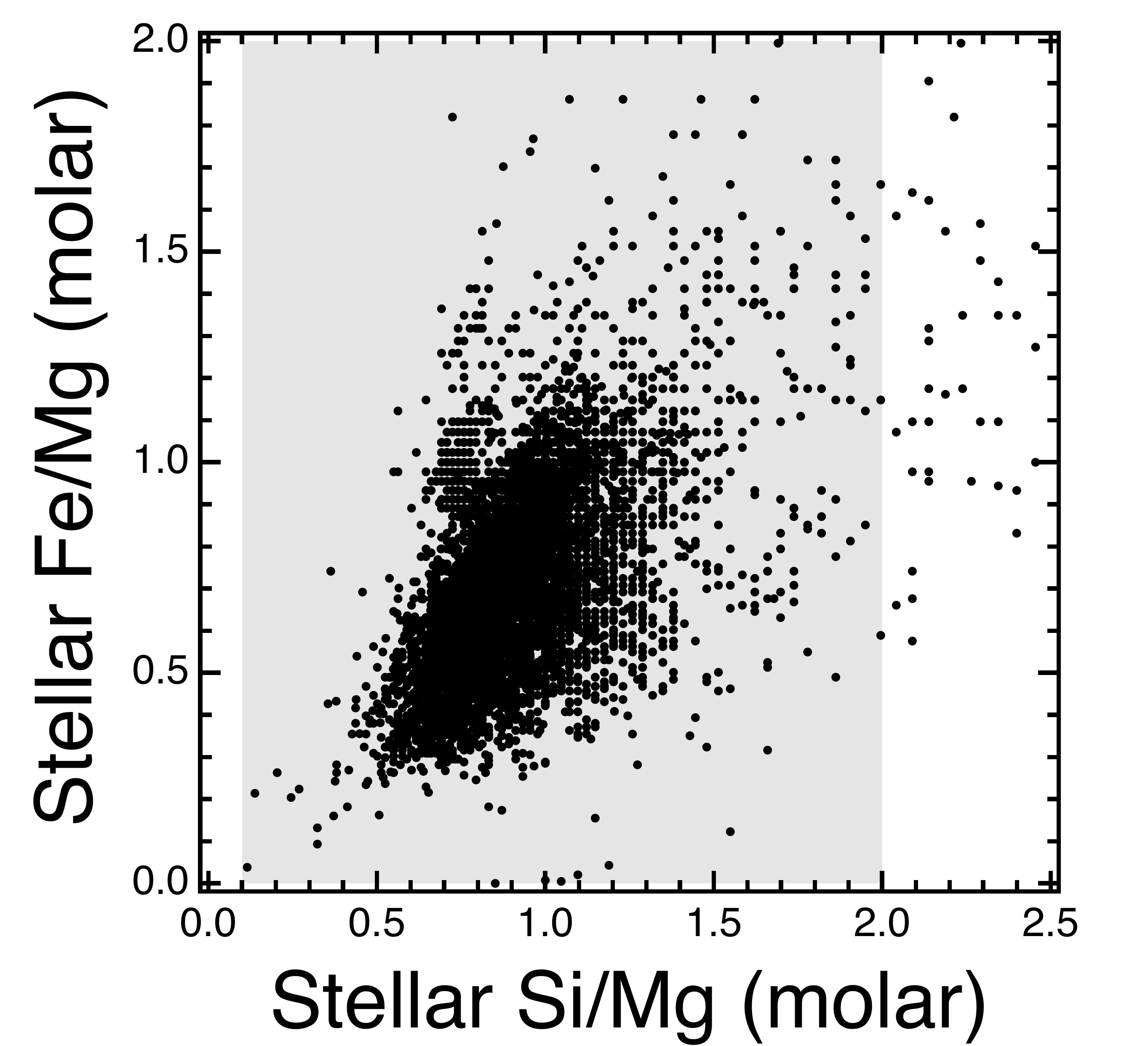}
    \caption{Molar Fe/Mg and Si/Mg from the Hypatia Catalog \citep{Hink14}. The range of Si/Mg compositions represented by the ExoPlex pre-made grids is shown as a gray band (see \S\ref{sec:Mantle}). }
    \label{fig:abund_hist}
\end{figure}

We adopt the Hypatia Catalog \citep{Hink14} to quantify the range of stellar, and thus potential bulk planet, compositions.
The Hypatia Catalog\footnote{\url{www.hypatiacatalog.com}} provides a standardized composite set of stellar abundance data that has been compiled from the literature and renormalized to the same solar abundance scale for stars near to the Sun (within 500 pc). 
When multiple literature sources report abundance measurements for the same element for the same star, that defines the \textit{spread} or the difference between the maximum and minimum measurement. 
The \textit{spread} provides a reasonable determination of how well an element within a star is truly measured---that is, the elemental precision---given the variety of telescopes, line lists, models, techniques, etc., employed by the stellar abundance community. 
Taking into account this spread and converting the stellar abundances to molar ratios \citep{Hink22}, we estimate the independent range of molar ratios Fe/Mg = $0.71 \pm 0.18$ and Si/Mg = $0.79 \pm 0.19$ (Figure \ref{fig:abund_hist}), however, these abundances correlated with each other.

With stellar abundances in hand, we can now explore the effects of this range of bulk compositions on exoplanet mass and radius across different mantle and core chemistries. 

\section{Effects of core mass fraction on planetary mass and radius}
\label{sec:baseline}

\begin{figure}[t]
    \centering
    \includegraphics[width=0.75\linewidth]{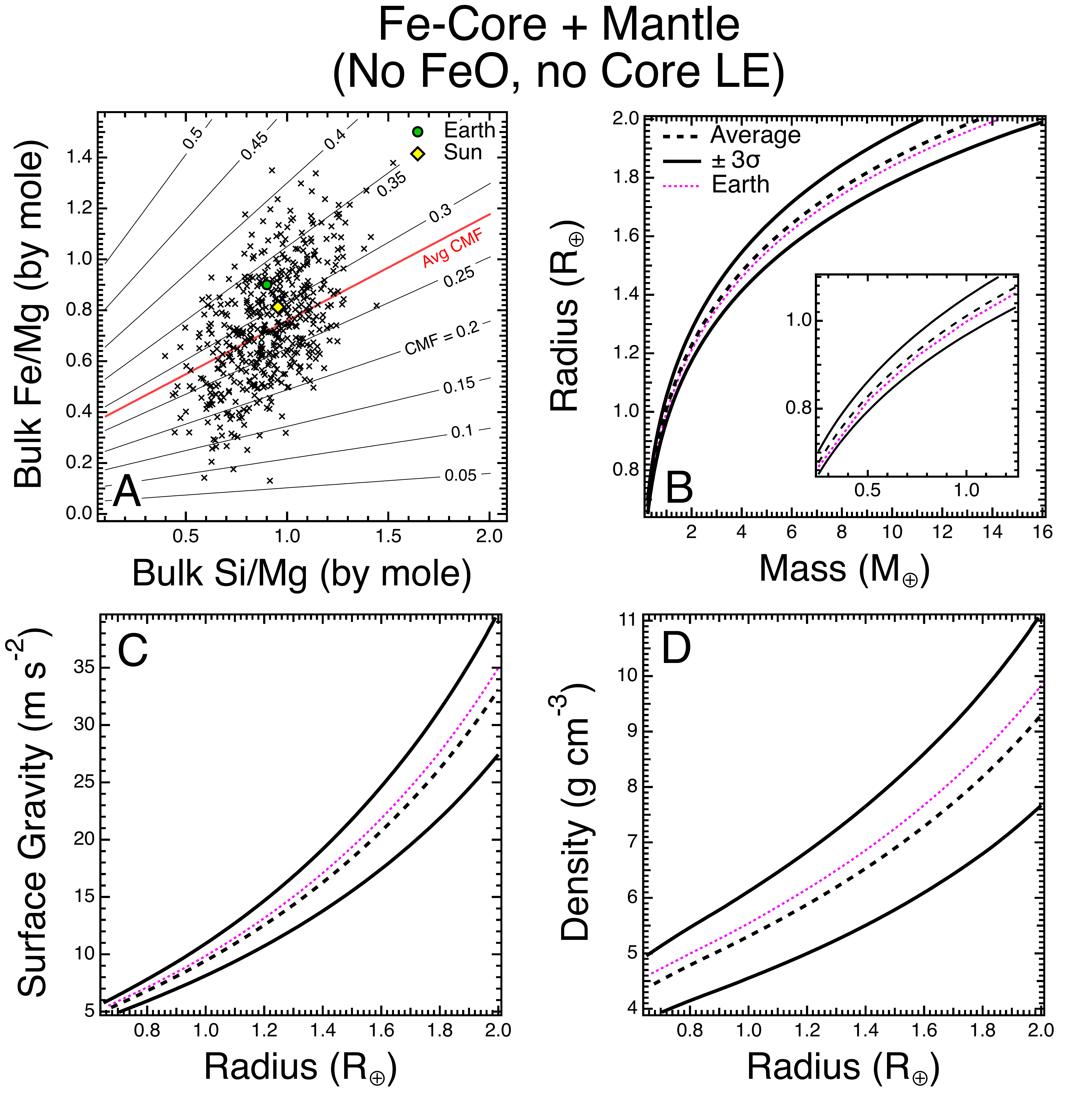}
    \caption{\textbf{A}: Contours of exoplanet core mass fraction (CMF) as a function of bulk Fe/Mg and Si/Mg, assuming all Fe remains in the core, there are no light elements (LE) in the core, and mantle Ca/Mg and Al/Mg are Earth-like. Markers represent the 500 abundances randomly sampled from the abundance values in Figure \ref{fig:abund_hist}. Red curve is the average CMF for this distribution. \textbf{B, C, D}: Planet radius as a function of input mass (B), and surface gravity (C), and bulk density (D), as a function of planet radius, for each composition from A. In each subfigure, the output is fit as a Gaussian, and curves denoting the mean (black dashed) and 3$\sigma$ (99.7\%; black solid) bounds are plotted, along with a curve for an Earth-like composition (fuchsia dashed).}
    \label{fig:baseline}
\end{figure}

A planet's bulk composition affects its bulk density. 
For those planets without significant surface volatiles, the primary compositional factor affecting a planet's mass for a given radius is the relative size of its core, as captured by its core mass fraction (CMF). 
The relative sizes of a planet's core and mantle are complex functions of the total amount of Fe present relative to the other rock-forming elements, and the oxidation state of the core and mantle. 
Starting simply, if all Fe is present in the core, a planet's CMF is simply the mass of the all Fe in the planet divided by the mass of the planet: 
\begin{equation}
    \rm{CMF} = \frac{\rm{(Fe/Mg)}*\mu_{\rm{Fe}}}{\rm{(Fe/Mg)}*\mu_{\rm{Fe}} + \sum_{\mathrm{X = Mg, Si, Al, Ca}}\left(\textit{X}/\mathrm{Mg}\right)*\left(\mu_{\textit{X}} + n_{\textit{X}}*\mu_{\mathrm{O}}\right)}
    \label{eq:cmf}
\end{equation}
where $\mu_{x}$ is the molar weight and ${\rm n}_{X}$ is the number of O atoms in the oxide of element $X$. 
We find it useful to define 
\begin{equation}
\bar{\mu} = \sum_{X = {\rm Al},{\rm Ca},{\rm Si},{\rm Mg}} \, \left( \frac{X}{\rm Mg} \right) * \left( \mu_{\rm X} + {\rm n}_{\rm X} * \mu_{\rm O} \right),
\end{equation}
where ${\rm n}_{\rm X}\!\!=\!\!1$ for Mg and Ca, 1.5 for Al, and 2 for Si.
In terms of $\bar{\mu}$, the core mass fraction is ${\rm CMF} = ({\rm Fe}/{\rm Mg}) \mu_{\rm Fe} /$ $[ ({\rm Fe}/{\rm Mg}) \mu_{\rm Fe} + \bar{\mu} ]$ in the simple case of iron core and oxide mantle.
Although the abundances of Ca and Al matter, through $\bar{\mu}$, CMF is primarily a function of both Fe/Mg and Si/Mg (Figure~\ref{fig:baseline}A), as Al and Ca oxides typically comprise only $\sim 5\%$ of a planet's mass. 
At low Fe/Mg, CMF is relatively insensitive to Si/Mg, primarily due to the lack of total Fe. 
As Fe/Mg increases, a planet's CMF depends more on Si/Mg, primarily due to differences in molar weight between the prominent oxides MgO (40.3 g/mol) and SiO$_2$ (60.1 g/mol): as Si/Mg increases, the relative mass of the mantle increases as more SiO$_2$ is added, thus lowering a planet's CMF. 
To quantify the range of likely compositions present within rocky exoplanets, we must first look to the range outlined by stellar abundance data. 

\begin{figure}
    \centering
    \includegraphics[width=.9\linewidth]{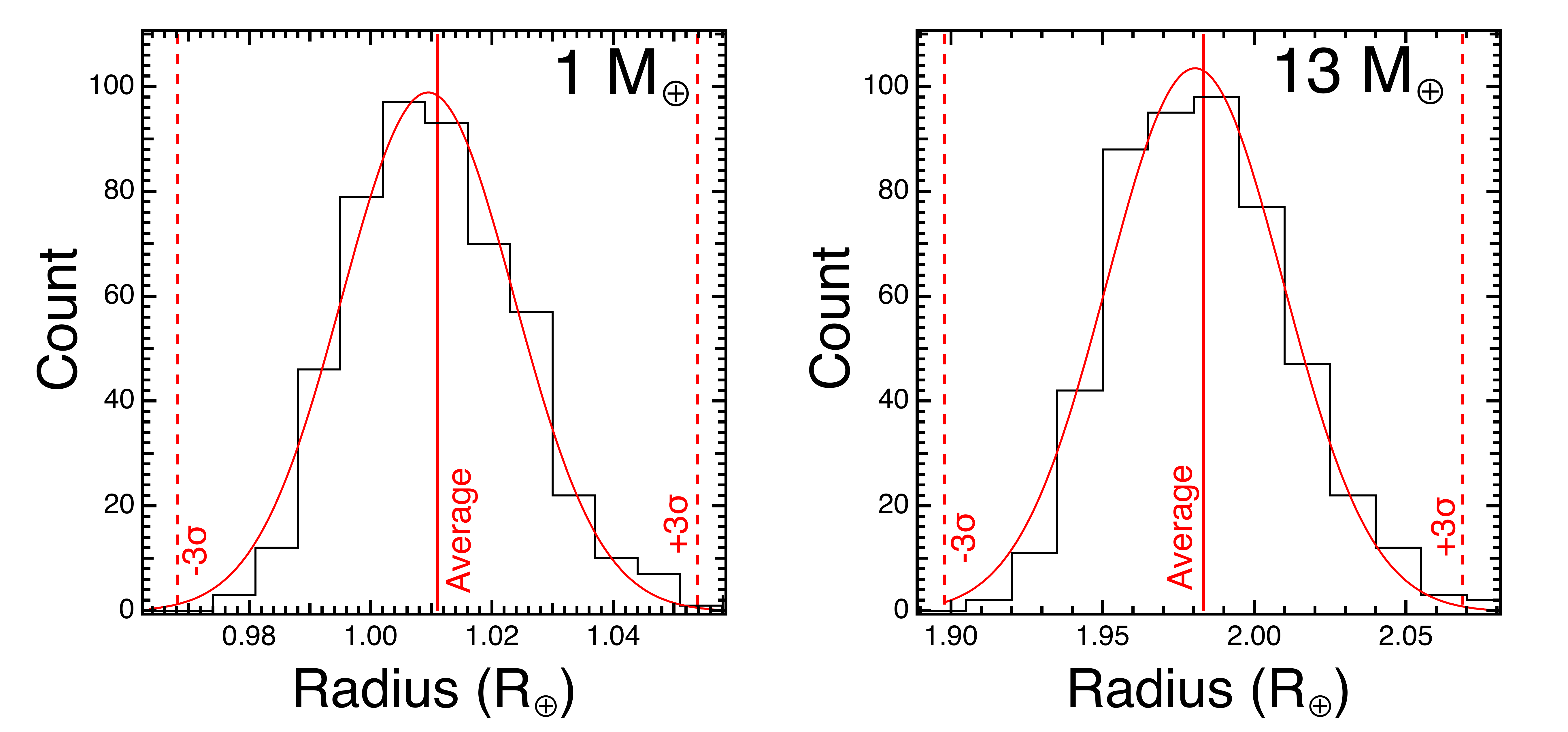}
    \caption{Resulting histograms (black) and Gaussian best-fit (red) of calculated planetary radius for 1 (left) and 13 (right) M$_\oplus$ planets using 500 randomly sampled stellar compositions within the Hypatia Catalog (Figure \ref{fig:baseline}A). These models assume all Fe is present within the core. Average and $\pm3\sigma$ radii from the Gaussian fit are included for reference.}
    \label{fig:distro}
\end{figure}

To explore the effects of Fe/Mg and Si/Mg (and CMF) on planet radius, we randomly sampled 500 Fe/Mg and Si/Mg pairs (Figure~\ref{fig:baseline}A) assuming a multi-variate normal distribution in order to account for the correlation in both abundance ratios in the stellar abundance data set (Figure \ref{fig:abund_hist}). 
These sampled compositional points yield a distribution of planetary CMFs, with an average CMF of 0.29.
This is equivalent to setting the average Fe/Mg = 0.71, and adopting $\bar{\mu} = 97.2$ g/mol.  
For comparison, a planet with Earth-like or solar Fe/Mg and Si/Mg abundances have CMFs of 0.33 and 0.30, respectively. 
We then ran ExoPlex to determine the radius for planet masses between 0.25 and 16 M$_\oplus$ for each of these 500 abundance pairs (Figure~\ref{fig:baseline}B, C, D).
We then calculated the average and 99.7\% ($3\sigma)$ bounds of the resulting radius distribution for a given input mass by fitting to a Gaussian (Figure \ref{fig:distro}). 
At 1 M$_\oplus$, 99.7\% of all models fall between $0.97$ and $1.05 \, R_{\oplus}$, and between $1.9$ and $2.07 \, R_{\oplus}$ at 13 M$_\oplus$ (Figure \ref{fig:distro}). 
\citet{Romy21} found that when taking into account all observational uncertainties, a measurements of a planet's surface gravity has the highest achievable precision, followed by its density and then mass. 
Using the mass as input and outputting radius, we calculate that 99.7\% of our surface gravities fall between 8.1 and $10.9 \, {\rm m} \, {\rm s}^{-2}$ at $1 \, R_{\oplus}$, and $27$ to $39 \, {\rm m} \, {\rm s}^{-2}$ at radius $\approx 2 \, R_{\oplus}$ (Figure~\ref{fig:baseline}C). 
We find bulk planet density varies between $4.6$ and $6.1 \, {\rm g} \, {\rm cm}^{-3}$ at radius $1 \, R_{\oplus}$, and between $\approx 7.7$ and $11 \, {\rm g} \, {\rm cm}^{-3}$ at radius $\approx 2 \, R_{\oplus}$ (Figure~\ref{fig:baseline}D). 

\begin{figure}
    \centering
    \includegraphics[width=\linewidth]{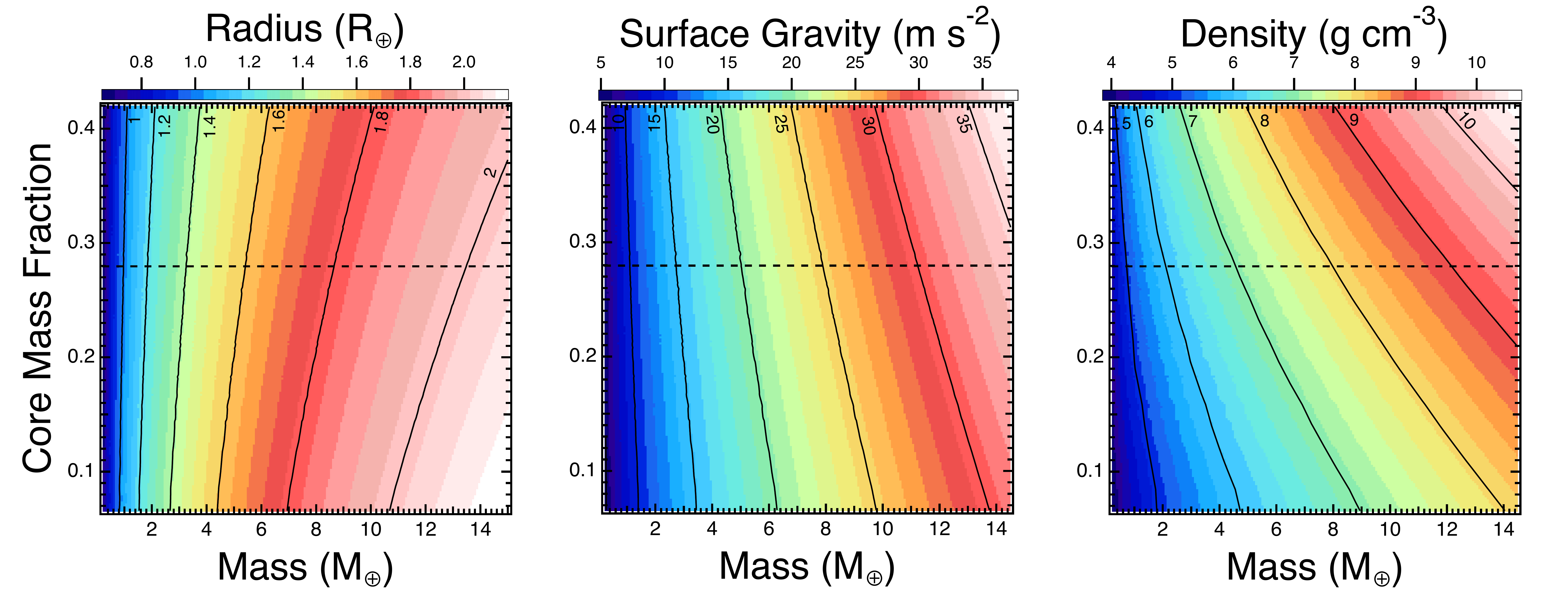}
    \caption{Interpolated contours of planet radius (left), surface gravity (center) and bulk density (right) as functions of planet mass and core mass fraction, using data taken from Figure~\ref{fig:baseline}. 
    The average CMF ($= 0.28)$ is included as a dashed line.
    }
    \label{fig:cmf_baseline}
\end{figure}

For this simple model of planet structure, we find that planet radius is relatively insensitive to CMF for planets less massive than $\sim 4 \, M_{\oplus}$ (Figure~\ref{fig:cmf_baseline}).
If a planet has given mass $M_{\rm p}$ and core mass fraction CMF, with mantle density $\rho_{\rm m}$ and core density $\rho_{\rm core}$, it is simple to show that 
\begin{equation}
R_{\rm p} = \left( \frac{3 \, M_{\rm p}}{4\pi \, \rho_{\rm core}} \right)^{1/3} \, \left[ {\rm CMF} +` \left( \frac{\rho_{\rm core}}{\rho_{\rm m}} \right) \, \left( 1 - {\rm CMF} \right) \right]^{1/3}.
\label{eq:radius}
\end{equation}
At low pressures, the core and mantle densities can be assumed fixed.
Taking the partial derivatives of Equation~\ref{eq:radius} with respect to $M_{\rm p}$ and CMF, the fractional change in $R_{\rm p}$ is found to be $\partial \ln R_{\rm p} / \partial \ln M_{\rm p} \approx +0.33$, but $\partial \ln R_{\rm p} / \partial \ln {\rm CMF} \approx -0.08$, meaning $R_{\rm p}$ is much more sensitive to mass than CMF. 
As planet masses increase such that central pressures are well above the bulk modulus of liquid \citep[$\approx 110$ GPa;][]{Ander94} or solid Fe \citep[$\approx 160$ GPa;][]{Dewaele17}, the core density increases significantly; the increase in mantle density, even when it is compressed, is not as significant as the compression in the core, because the bulk modulus of minerals like bridgmanite are greater \citep[$\approx 250$ GPa;][]{Stix11}.
Core-mantle boundary pressures significantly above 100 GPa require the planets being a few $M_{\oplus}$ in mass \citep{Unter19}. 
As the $\rho_{\rm core} / \rho_{\rm m}$ ratio increases, $\partial \ln R_{\rm p} / \partial \ln {\rm CMF} \approx -0.16$, showing that $R_{\rm p}$ becomes half as sensitive to CMF as it is to $M_{\rm p}$.
Likewise, bulk density must follow 
\begin{equation}
\bar{\rho} = \rho_{\rm m} \, \left[ 1 - \left( 1 - \frac{\rho_{\rm m}}{\rho_{\rm core}} \right) \, {\rm CMF} \right]^{-1}.
\end{equation}
These trends, and the trends in surface gravity and bulk density, are understood in terms of the liquid iron core being more compressible than the mantle, and are illustrated in  Figure~\ref{fig:cmf_baseline}.

This simple model of determining a planet's CMF ignores many chemical processes that can alter a rocky exoplanet's structure, mineralogy and distribution of elements within its interior. 
It does, however, provide a baseline range of observable values that will allow us to compare how geochemical and geophysical processes can affect a planet's CMF, radius, surface gravity and density as mass and bulk composition change. 
Below we explore various other possible mantle/core geochemistries and volatile contents to quantify these differences. 

\section{Effects of mantle FeO content on planetary mass and radius}
\label{sec:FeO}
Fe can be present within the core in its metallic phase (Fe), as well as in the mantle in its oxidized form, FeO, usually in solid-state solutions like $({\rm Mg,Fe})_{2}{\rm SiO}_{4}$, $({\rm Mg,Fe}){\rm SiO}_{3}$, or $({\rm Mg,Fe}){\rm O}$. 
The mantles of the Earth and Mars are $\sim 8$ and 18 weight percent FeO, respectively \citep{McD03,Wanke1994}. 
For the Earth, this means $\sim 14\%$ of all Fe atoms reside in the mantle as FeO \citep{McD03}. 
The exact distribution of Fe between the core and mantle is a complex function of the pressure and temperature during core-mantle equilibration after planet formation, and the overall oxidation state of the planet \citep[e.g.,][]{Schaefer2017, Fisch20, Unterborn20}. 

\begin{figure}
    \centering
    \includegraphics[width=0.75\linewidth]{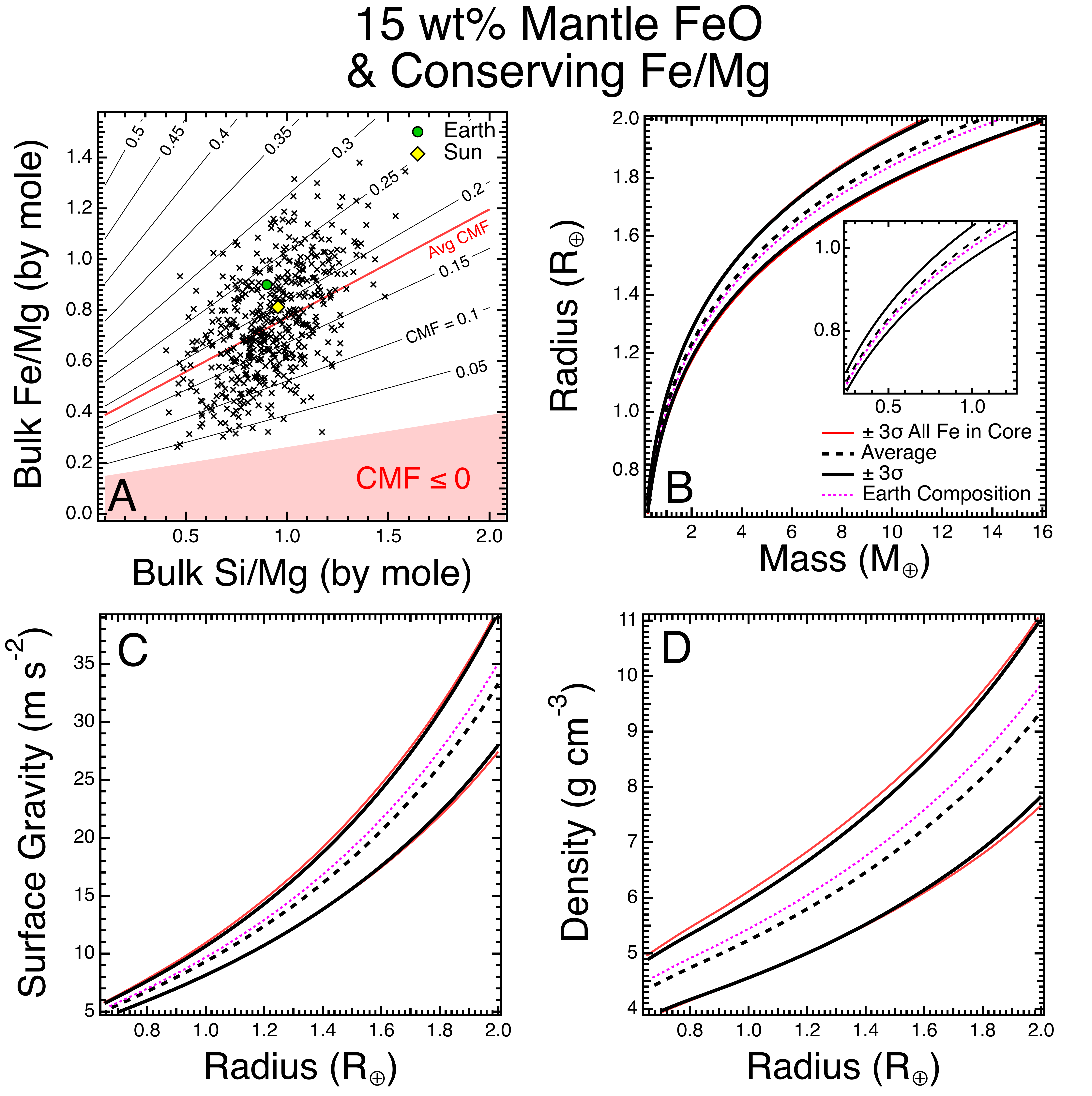}
    \caption{\textbf{A}: Contours of exoplanet core mass fraction (CMF) as a function of bulk Fe/Mg and Si/Mg for planets containing 15 wt\% mantle FeO. CMF is determined following equation \ref{eq:cmf_feo}. Those compositions that would produce CMF $<0$ (insufficient iron to supply the mantle FeO) are shown in red. Markers represent the 500 abundance pairs randomly sampled from the abundance values in Figure \ref{fig:abund_hist}. Red curve is the average CMF for this distribution. \textbf{B, C, D}: Planet radius as a function of input mass (B), and surface gravity (C), and bulk density (D), as a function of planet radius (black lines), for each composition from A. In each subfigure, the output is fit as a Gaussian, and curves denoting the mean (dashed black) and 3$\sigma$ (99.7\%, solid black) bounds are plotted, along with a curve for an Earth-like composition (fuchsia dashed). The $\pm3\sigma$ values from the baseline (0 wt\% FeO) model from Figure~\ref{fig:baseline} are shown as red lines.}
    \label{fig:FeO_diffs}
\end{figure}

In ExoPlex, the user defines the molar abundance ratios and the weight fraction of FeO in the mantle, ${\rm wt}\%{\rm FeO}^{\rm Man}$.
The fraction of Fe atoms (by mole) that are in the mantle is found to be 
\begin{equation}
\frac{\%{\rm Fe}^{\rm Man}}{100} = \frac{ w_{\rm FeO} }{1 - w_{\rm FeO} } \, \frac{ \bar{\mu} }{ ({\rm Fe}/{\rm Mg}) },
\end{equation}
where $w_{\rm FeO} = ({\rm wt}\%{\rm FeO}^{\rm Man} / 100)$.
The core mass fraction is 
\begin{equation}
{\rm CMF} = 
\frac{ \left[ 
\left( \frac{\rm Fe}{\rm Mg} \right) \, \mu_{\rm Fe} -  \frac{\bar{\mu}}{\mu_{\rm FeO}} \, \frac{ w_{\rm FeO} \, \mu_{\rm Fe} }{ 1 - w_{\rm FeO} }
\right] }
{ \left[ \left( \frac{\rm Fe}{\rm Mg} \right) \, \mu_{\rm Fe} + \frac{\bar{\mu}}{\mu_{\rm FeO}} \, \frac{ \mu_{\rm FeO} - w_{\rm FeO} \, \mu_{\rm Fe} }{ 1 - w_{\rm FeO} }
\right] }
\label{eq:cmf_feo}
\end{equation}
(assuming no light elements in the core).
We can also write the mantle mass fraction as
\begin{equation}
1 - {\rm CMF} = \left\{ \left( 1 + \left( \frac{\rm Fe}{\rm Mg} \right) \, \frac{ \mu_{\rm Fe} }{ \bar{\mu} } \right) - w_{\rm FeO} \, \left( \left( \frac{\rm Fe}{\rm Mg} \right) \, \frac{\mu_{\rm Fe} }{ \bar{\mu} } + \frac{ \mu_{\rm Fe} }{ \mu_{\rm FeO} } \right) \right\}^{-1}.
\end{equation}
As the bulk Fe/Mg ratio is a fixed input to ExoPlex, the total iron content of the planet is fixed.
The mantle incorporates FeO in the mantle at the expense of Fe in the core, and there is a minimum bulk Fe/Mg to supply the mantle FeO content, $({\rm Fe} / {\rm Mg})_{\rm min} = [w_{\rm FeO}/(1-w_{\rm FeO})] \, \bar{\mu} / (\mu_{\rm FeO} )$.
For 15 wt\% FeO in the mantle, $({\rm Fe}/{\rm Mg})_{\rm min} = 0.15$ for Si/Mg = 0.5, increasing with increasing Si/Mg (Figure~\ref{fig:FeO_diffs}A).
In general, the CMF decreases with increasing ${\rm wt}\%{\rm FeO}^{\rm Man}$, and a planet with 15 wt\% FeO in the mantle has a smaller core than one with no FeO in the mantle.

Using ExoPlex, we again calculate the planet radius, surface gravity, and bulk density, following the same methodology as Figure~\ref{fig:baseline}, and sampling 500 new bulk Fe/Mg and Si/Mg pairs; but now we assume 15 wt\% FeO in the mantle 
(Figure~\ref{fig:FeO_diffs}A).
This represents the simplest model for the creation of mantle FeO, where Fe simply arrives in its oxidized form while still preserving the bulk Fe/Mg ratio. 
This model, however, does not conserve the total fraction of O within the planet. 
As mantle FeO content increases, the relative amount of O relative to the other cations increases as well. 
We explore more complex pathways to FeO production in \S\ref{sec:core_LE}.
We calculate an average CMF of 0.19, about 30\% lower than the case with no mantle FeO shown in Figure~\ref{fig:baseline}A, which had CMF of 0.29.
This CMF is consistent with the value predicted by Equation~\ref{eq:cmf_feo} assuming Fe/Mg = 0.71 and $\bar{\mu} = 97.2$ g/mol. 

In Figures~\ref{fig:FeO_diffs}B, C, D, we plot the planet radius, surface gravity, and bulk density, as in Figure~\ref{fig:baseline}. 
At 1 M$_\oplus$, 99.7\% of all models fall between $0.97$ and $1.05 \, R_{\oplus}$, and between $1.9$ and $2.06 \, R_{\oplus}$ at 13 M$_\oplus$ (Figure~\ref{fig:FeO_diffs}B).  
Using the mass as input and outputting radius, we calculate that 99.7\% of our surface gravities fall between 8.1 and $10.7 \, {\rm m} \, {\rm s}^{-2}$ at $1 \, R_{\oplus}$, and $28$ to $39 \, {\rm m} \, {\rm s}^{-2}$ at radius $\approx 2 \, R_{\oplus}$ (Figure~\ref{fig:FeO_diffs}C). 
We find bulk planet density varies between $4.6$ and $5.7 \, {\rm g} \, {\rm cm}^{-3}$ at radius $1 \, R_{\oplus}$, and between $\approx 7.8$ and $11 \, {\rm g} \, {\rm cm}^{-3}$ at radius $\approx 2 \, R_{\oplus}$ (Figure~\ref{fig:FeO_diffs}D). 
Despite the 15 wt\% FeO mantle model having a lower CMF, compared to the baseline FeO-free model with all Fe in the core, we find only a small reduction in bulk density; in fact, the $-3\sigma$ curves for each are practically indistinguishable. 
Although the core shrinks as more Fe partitions into the mantle as FeO, the density of the mantle also increases, due to FeO having a higher molar weight (71.8 g/mol) than MgO (40.3 g/mol) and SiO$_2$ (60.1 g/mol). 
Additionally, the FeO-bearing silicates within ExoPlex have compressibilities similar or slightly smaller than their MgO-bearing counterparts in the mantle \citep{Stix11}, leading to comparable or slightly higher mantle density than the FeO-free case.
The increase in mantle density, therefore, almost exactly cancels out the effects of the planet having a smaller central Fe-core, at least for 15 wt\% Fe. 

To examine the effect of mantle FeO content on planet size more generally, we reran the highest and lowest CMF models of the 500 random Si/Mg and Fe/Mg samplings from Figure~\ref{fig:FeO_diffs}A at planet masses of 1 and 9 M$_\oplus$.
These compositions are Si/Mg = 0.94 and Fe/Mg = 0.27 for the low-CMF model, and Si/Mg = 0.32 and Fe/Mg = 1.17 for the high-CMF case. 
We then varied the mantle FeO content from 0 wt\% (all Fe in core) to the point where all Fe is present in the mantle and the planet has no central core. 
For those interior models with $\leq 20$ wt\% mantle FeO, ExoPlex pre-defined grids were used. 
For higher FeO contents, we directly calculated new grids of thermoelastic parameters using \texttt{Perple\_X}.  

\begin{figure}
    \centering
    \includegraphics[width=0.75\linewidth]{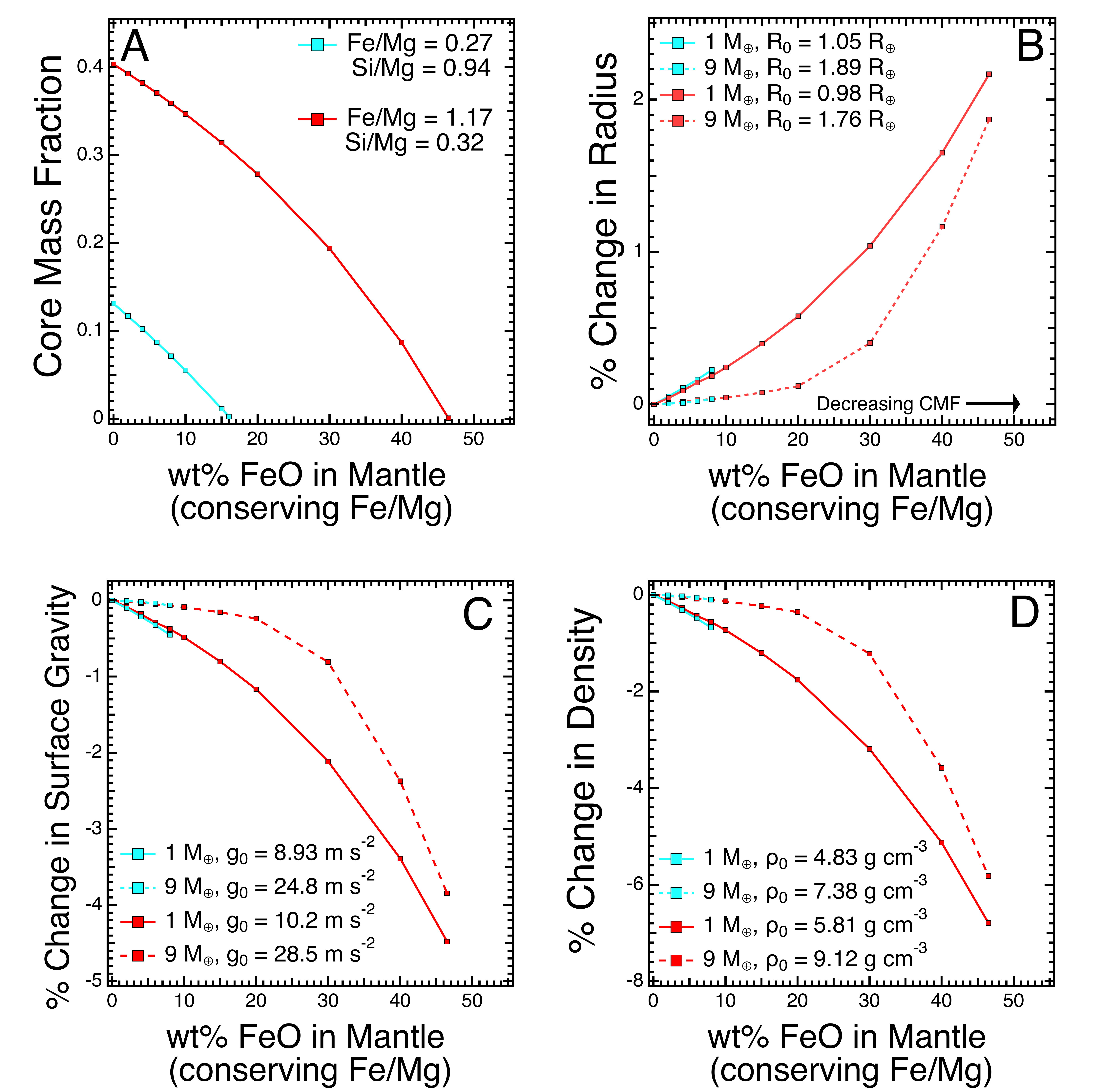}
    \caption{Core mass fraction (\textbf{A}) and fractional change in planet radius (\textbf{B}), surface gravity (\textbf{C}) and bulk density (\textbf{D}) for our low-CMF (teal) and high-CMF (red) models in a 1 M$_\oplus$ (solid) and 9 M$_\oplus$ (dashed) planet, as the mass fraction of FeO in the mantle increases. The exact Fe/Mg and Si/Mg values for the low and high CMF cases are noted in the figure. }
    \label{fig:feo_per}
\end{figure}

We estimate that a planet becomes coreless when the mantle contains $\sim$15 and $\sim$50 wt\% FeO for the low- and high-CMF cases, respectively (Figure \ref{fig:feo_per}A).
The difference in these values is due entirely to the low-CMF case simply having fewer moles of Fe available to put into the mantle as FeO, due to its low Fe/Mg. 
We find that planet radius increases as mantle FeO content increases, by at most 0.2\% and 2.2\% for the low- and high-CMF cases, respectively, when bulk Fe/Mg is conserved (Figure \ref{fig:feo_per}B).
This translates to maximum decreases of $\sim 0.45 \%$ and 4.5\% of in surface gravity, and of 0.7\% and 6.8\% in bulk density, for the high- and low-CMF cases, respectively (Figure \ref{fig:feo_per}C, D). 
These maximum changes, however, are when a planet lacks a central Fe-core almost entirely.
At Earth-like mantle FeO contents ($\sim$8 wt\%), our estimated changes in planet radius, surface gravity and bulk density are all $<$1\%. 

This simple stoichiometric treatment has not specified the origin of the FeO.
Carbonaceous chondrites are an example of planetesimals that are hydrated, with FeO contents exceeding tens of wt\% \citep{WassonKallemeyn1988}, so some of the iron accreted by a planet could be accreted as FeO.
It is also possible the FeO was produced from metallic ${\rm Fe}^{0}$ within the planet by oxidation/reduction reactions like the following:
\begin{equation}
    \rm{Fe} + \frac{1}{2}\rm{O_{2}} \longrightarrow \rm{FeO}
\end{equation}
or
\begin{equation}
    \rm{Fe} + {\rm H}_{2}{\rm O} \longrightarrow \rm{FeO} +{\rm H}_{2}.
\end{equation}
The first reaction is unrealistic, as there is no source of free oxygen.
In contrast, the second reaction (similar to serpentinization) is likely, as water is a plentiful source of oxygen, and the hydrogen can escape if generated near the surface.
A third reaction is also possible in chemically reducing mantles, in which Fe reacts with oxygen in ${\rm SiO}_{2}$, forming Si metal, as explored below in \S~\ref{sec:core_LE}.
While accretion of FeO is very plausible, so is production of FeO within the planet, and evidence points to some amount of core-mantle equilibration occurring during a planet's early evolution \citep[e.g.,][]{Schaefer2015}. 
ExoPlex is agnostic on the issue of whether FeO is accreted or produced within a planet.  
As described above, ExoPlex does not conserve the total amount of oxygen present within the planet; instead it simply conserves the molar ratios of the planet-building elements (e.g., Fe/Mg), while providing as much O as is needed to oxidize the refractory elements within the mantle and produce the specified FeO content. 
In either scenario, if the bulk Fe/Mg ratio in a planet is fixed, then there is a trade-off between the amounts of Fe in the core and FeO in the mantle. 


\section{Effects of core light elements on planetary mass and radius}
\label{sec:core_LE}

\subsection{Coupled FeO and core light element production}

Elements other than Fe are present within the Earth's core \citep{Birch52}. 
The most abundant of these elements, at $\approx5$ wt\%, is Ni \citep{McD03} and is not currently included in our treatment of the core (but is expected to be included in a future update), as its effect on the EoS of iron is negligible, given that it has roughly the same molar mass as Fe and comparable thermoelastic parameters \citep{li2002,Wicks18}. 
Ni also partitions into NiO in the Earth's mantle almost as efficiently as Fe partitions into FeO. 
To include Ni in our calculations, it would be sufficient to increase the Fe abundance by 5.7\% \citep{Lodd03}, equivalent to 0.024 dex, or using an average Fe/Mg ratio of 0.75 instead of 0.71. 

Elements of lower molar weight---especially H, C, O Si and S---also can alloy with Fe, and would do so in sufficient abundance to lower the density of the core at a given pressure and temperature, compared to a pure Fe core \citep{Birch52, Unter16, SchlichtingYoung22}.  
Earth's core, for example, contains $\sim$ 8 wt\% other elements; however, the exact identity of the element(s) responsible for the lower density (``density deficit'') of Earth's core is still under debate \citep[e.g.,][]{McD03,Badro15,Li_LE_19,Fisch20}. 

These light elements likely alloyed with the Fe metal during partitioning between metal and silicates in the proto-mantle, during the magma ocean phase prior to the gravitational segregation of the core. 
During equilibration, light elements enter metal bound for the core, typically by oxidation/reduction reactions. 
A simple yet important example of this is the following reaction that can place Si into the core and FeO in the mantle: 
\begin{equation}
    3 \, \rm{Fe} +\rm{SiO_{2}} \longrightarrow 2 \, \rm{FeO} \;[\rm{to \;mantle}] + \rm{FeSi}\; [\rm{to\; core}]. 
    \label{eq:Si}
\end{equation}
In this reaction, two moles of Fe are oxidized using the O available from SiO$_2$, while the Si is reduced and alloys with the metallic Fe that segregates into the core.
This coupled oxidation/reduction reaction therefore has a two-fold effect on the planet's interior density structure: it lowers the density of the core due to Si's lower molar weight, while also placing FeO into the mantle, increasing its density and shrinking the core.

We consider reaction~\ref{eq:Si} to be representative of how many light elements enter the core, and explore the effects of Si in the core.
Leaving the mass fraction of Si in the core, $w_{\rm Si} = ({\rm wt}\%{\rm Si}^{\rm Core}/100)$, and the mass fraction of FeO in the mantle, $w_{\rm FeO} = ({\rm wt}\%{\rm FeO}^{\rm Man}/100)$, as free parameters, but keeping the planetary bulk abundances Fe/Mg and Si/Mg fixed, one can solve for quantities like the core mass fraction:
\begin{equation}
{\rm CMF} = 
\frac{
(1 - w_{\rm FeO}) \, \left( \frac{\rm Fe}{\rm Mg} \right) \, \mu_{\rm Fe} - w_{\rm FeO} \, \frac{\mu_{\rm Fe}}{\mu_{\rm FeO}} \, \bar{\mu}
}{
(1 - w_{\rm FeO}) \, \left( \frac{\rm Fe}{\rm Mg} \right) \, \mu_{\rm Fe} - w_{\rm FeO} \, \frac{\mu_{\rm Fe}}{\mu_{\rm FeO}} \, \bar{\mu}
+ (1 - w_{\rm Si}) \, \bar{\mu} - w_{\rm Si} \, \frac{\mu_{\rm SiO2}}{\mu_{\rm Si}} \, \left( \frac{\rm Fe}{\rm Mg} \right) \, \mu_{\rm Fe}.
}
\end{equation} 
The mantle mass fraction is 
\begin{equation}
1 - {\rm CMF} = \left\{ 
1 + \frac{\mu_{\rm Fe}}{\bar{\mu}} \, \left( \frac{\rm Fe}{\rm Mg} \right) \, 
\times
\frac{
\left[
1 - w_{\rm FeO} \, \left( 
1 + \frac{\bar{\mu}}{\mu_{\rm FeO}} \frac{1}{{\rm Fe}/{\rm Mg}}
\right)
\right]
}
{
\left[
1 - w_{\rm Si} \, \left( 
1 + \frac{\mu_{\rm SiO2}}{\mu_{\rm Si}} \frac{\mu_{\rm Fe}}{\bar{\mu}} \frac{\rm Fe}{{\rm Mg}}
\right)
\right]
}
\right\}^{-1}
\label{eq:cmfbig}
\end{equation}
In the limit that the mass fraction of Si in the core is 0wt\% ($w_{\rm Si} = 0$), and the mass fraction of FeO in the mantle is 0wt\% ($w_{\rm FeO} = 0$), ${\rm CMF} = \left[ 1 + (\bar{\mu}/\mu_{\rm Fe}) / ({\rm Fe}/{\rm Mg}) \right]^{-1}$.

If it is assumed that all of the FeO in the mantle derived from reaction~\ref{eq:Si}, then the following relationship between $w_{\rm Si}$ and $w_{\rm FeO}$ can be derived
\begin{equation}
w_{\rm Si} = w_{\rm FeO} \, \frac{
(\mu_{\rm Si} / 2 \mu_{\rm FeO}) }
{
\left( \frac{\rm Fe}{\rm Mg} \right) \, \frac{\mu_{\rm Fe}}{\bar{\mu}} \, \left[ (1 - w_{\rm FeO}) + w_{\rm FeO} \, \frac{\mu_{\rm SiO2}}{2 \mu_{\rm FeO}} \right] - w_{\rm FeO} \, \frac{ \mu_{\rm Fe} - \mu_{\rm Si}/2 }{ \mu_{\rm FeO} } 
}.
\label{eq:wsi}
\end{equation}
Substitution of Equation~\ref{eq:wsi} into Equation~\ref{eq:cmfbig} yields
\begin{equation}
1 - {\rm CMF} = \left\{ 
1 + \frac{\mu_{\rm Fe}}{\bar{\mu}} \, \left( \frac{\rm Fe}{\rm Mg} \right) 
-w_{\rm FeO} \, \left[ \frac{ \mu_{\rm Fe} - \mu_{\rm Si}/2 }{ \mu_{\rm FeO} } + \left( \frac{\rm Fe}{\rm Mg} \right) \, \frac{\mu_{\rm Fe}}{\bar{\mu}} \left( 1 - \frac{\mu_{\rm SiO2}}{2 \mu_{\rm FeO}} \right) \right] 
\right\}^{-1}.
\end{equation}
This also reduces to ${\rm CMF} = \left[ 1 + (\bar{\mu}/\mu_{\rm Fe}) / ({\rm Fe}/{\rm Mg}) \right]^{-1}$ in the limit that $w_{\rm FeO}$ (and $w_{\rm Si}$) vanish. 
This latter relationship makes evident that increasing the mass fraction of FeO in the mantle, while assuming it corresponds to the mass fraction of Si in the core, demands a decrease in CMF. 
That is, the effect of removing Fe from the core dominates over the effect of adding Si to the core via reaction~\ref{eq:Si}.
Another useful formula is the Si/Mg ratio in the mantle:
\begin{equation}
\left( \frac{\rm Si}{\rm Mg} \right)_{\rm mantle} = \left( \frac{\rm Si}{\rm Mg} \right) - \frac{ w_{\rm FeO} \, \bar{\mu} }{ 2 (1-w_{\rm FeO}) \mu_{\rm FeO} + w_{\rm FeO} \, \mu_{\rm SiO2} }.
\end{equation}
This is always somewhat smaller than the bulk Si/Mg ratio of the planet, since all Mg is assumed to stay in the mantle, even if some Si resides in the core. 
For a planet with Earth values Fe/Mg, Si/Mg, Al/Mg and Ca/Mg \cite{McD03} and an Earth-like mantle FeO content of 8 wt\%, this model predicts the core would contain $\sim$3.7 wt\% Si and a resulting mantle Si/Mg of $\sim$0.8.
This is slightly more than half the predicted value of $\sim$6 wt\% core Si, yet yields a mantle of nearly the same Si/Mg as predicted for the Earth of 0.8 \citep{McD03}.
This simple model, however, does not include the effects of other core light elements, or indeed other abundant planetary elements as well such as S, Ni and H.
Despite this, this model predicts a planet of almost exaclty 1 R$_\oplus$.

\begin{figure}
    \centering
    \includegraphics[width=0.75 \linewidth]{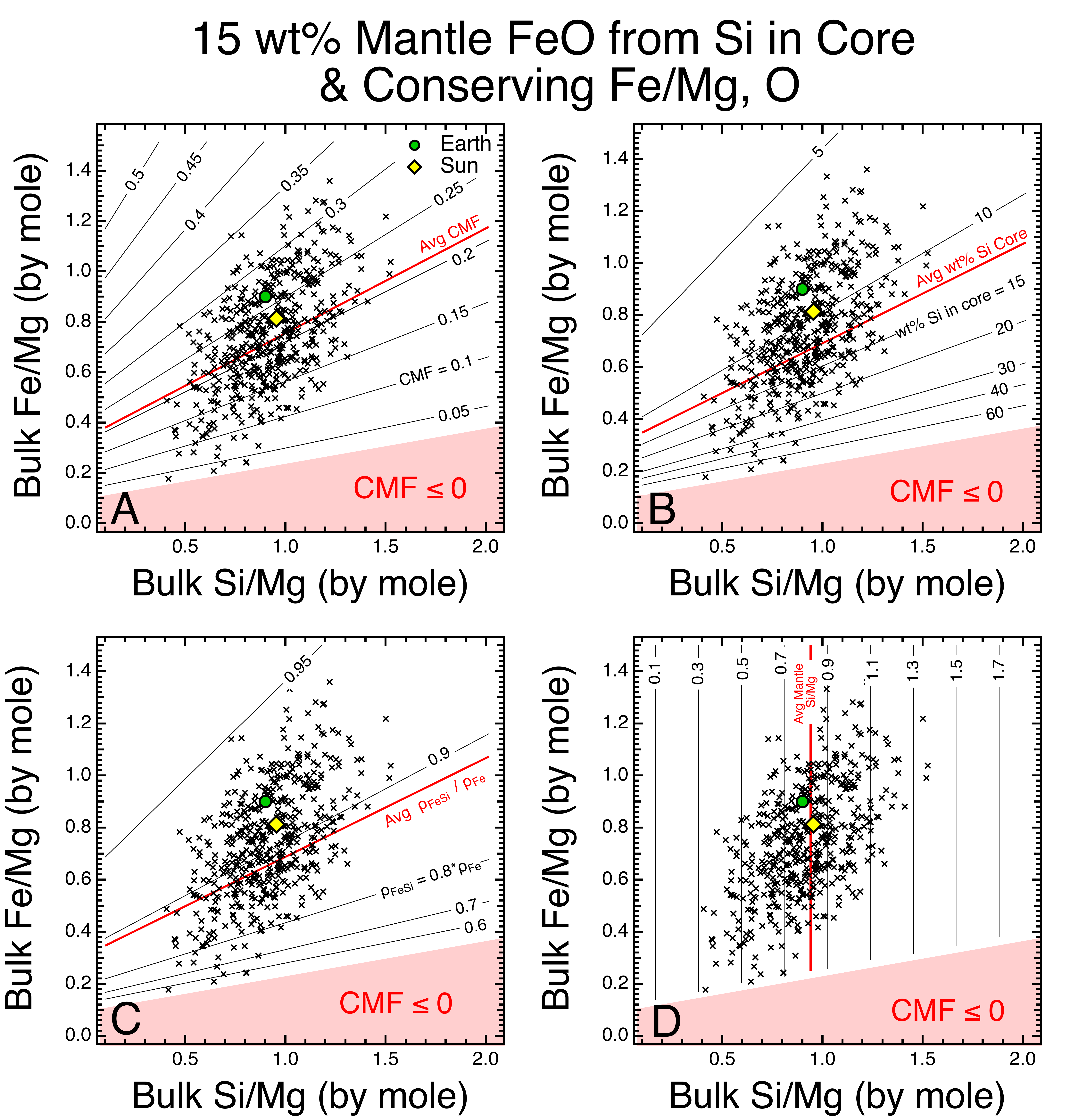}
    \caption{Contours of core mass fraction \textbf{(A)}, mass fraction of Si in the core \textbf{(B)}, approximate density of the core (relative to pure iron, \textbf{C}), and resulting mantle Si/Mg \textbf{(D)}, for planets containing 15 wt\% mantle FeO created through the reduction of Si (reaction~\ref{eq:Si}). For reference, the average values of each parameter using these 500 randomly sampled Fe/Mg and Si/Mg points are also shown (red).}
    \label{fig:comp_feo_wSi}
\end{figure}

To explore the effects of coupled mantle FeO and core Si production on a planet's derived properties, we again use ExoPlex to model planets between 0.25 and $\sim14 \, M_\oplus$ in mass, assuming the mantle contained 15 wt\% FeO due to the reduction of Si via reaction \ref{eq:Si}.  
We again randomly sampled 500 Fe/Mg and Si/Mg compositional pairs from the distribution of Figure~\ref{fig:abund_hist}, resampling those compositions that produced negative CMFs (Figure~\ref{fig:comp_feo_wSi}).
These random samples produced planets with CMFs between 0.03 and 0.35, with an average CMF of 0.22.
This is to be compared to the average CMF of 0.29 for the FeO-free case in (Figure~\ref{fig:baseline}), showing that putting Fe in the mantle (as FeO) while keeping Fe/Mg fixed lowers CMF.
This is to be compared as well to the case of planets with 15 wt\% FeO in their mantles, but no Si in their cores (Figure~\ref{fig:FeO_diffs}); these cases had average CMF of 0.19. 
The slight difference arises from the trade offs of putting Si in core at the expense of the mantle, and the associated non-conservation of oxygen.
These compositions also represent core silicon abundances of between $\sim$5 and 65 wt\% and mantle Si/Mg between 0.5 and 1.23 and density reductions in the core between $\sim5$ and 40\%  (Figure~\ref{fig:comp_feo_wSi}).

\begin{figure}
    \centering
    \includegraphics[width=\linewidth]{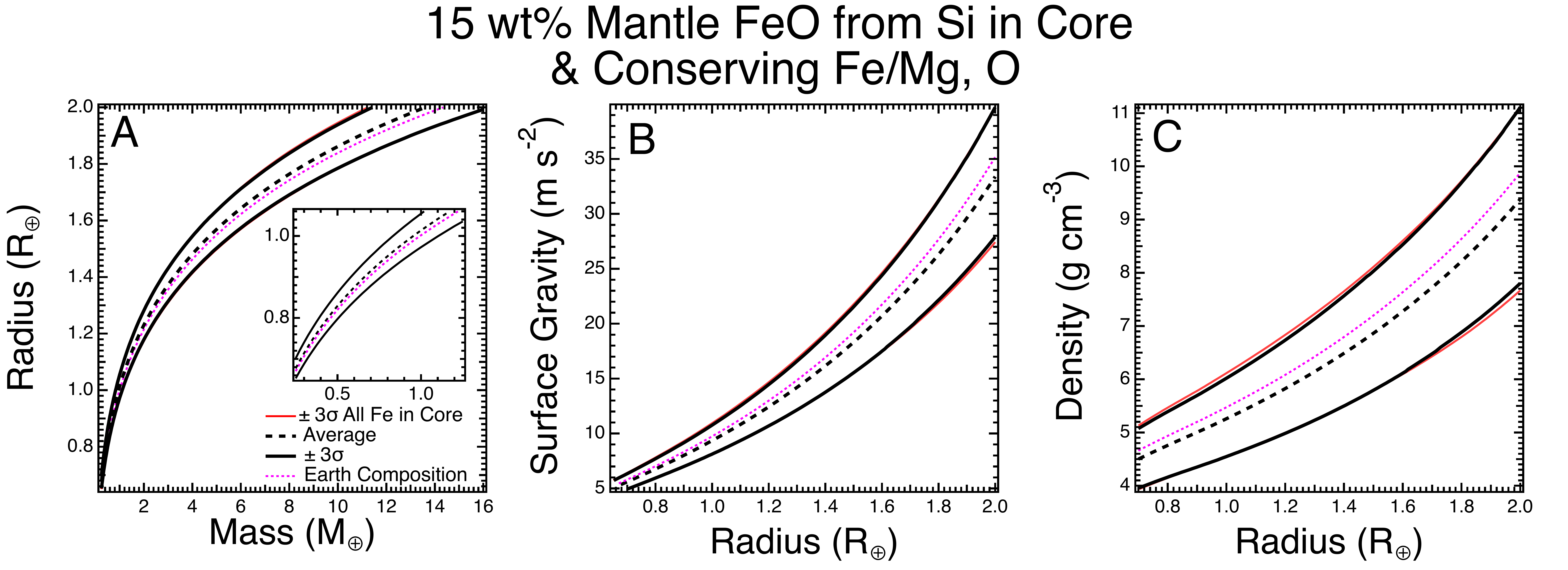}
    \caption{Planet radius (\textbf{A}), surface gravity (\textbf{B}) and bulk density (\textbf{C}), for planets containing 15 wt\% FeO created via Si entering the core (reaction \ref{eq:Si}). In each subfigure, the output for each mass is fit as a Gaussian, and curves denoting the mean (black dashed) and 3$\sigma$ (99.7\%; black solid) bounds are plotted, along with a curve for an Earth composition (fuchsia dashed). The $\pm3\sigma$ values from the baseline (0 wt\% FeO, no core light elements) model from Figure~\ref{fig:baseline} are shown as red lines.}
    \label{fig:LE_wSi_MR}
\end{figure}

\begin{figure}
    \centering
    \includegraphics[width=0.75\linewidth]{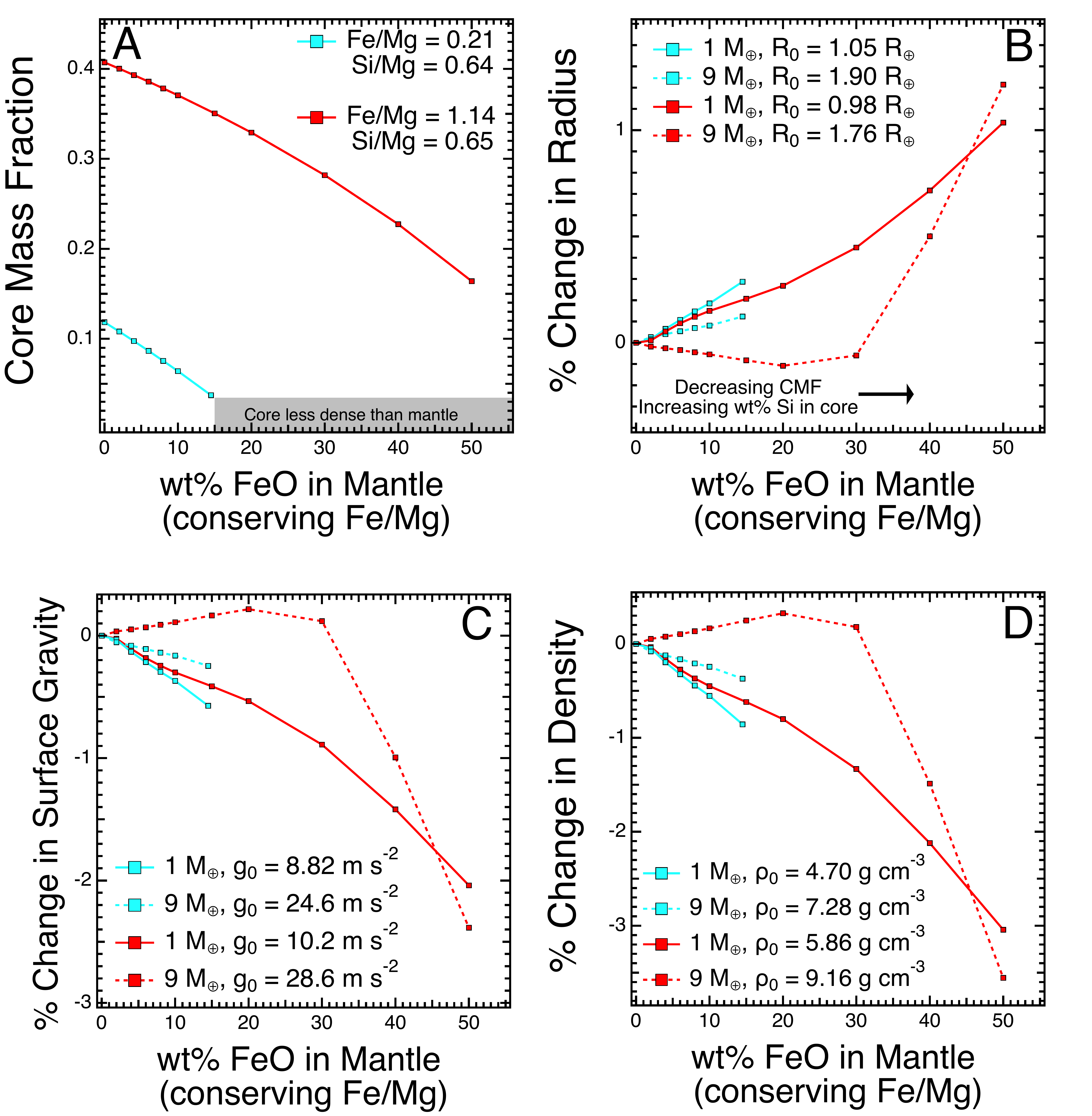}
    \caption{Core mass fraction (\textbf{A)}, percent change in planet radius (\textbf{B}), surface gravity (\textbf{C}) and density (\textbf{D}), as a function of mantle FeO content, for our lowest-CMF (teal) and highest-CMF (red) models in a $1 \, M_{\oplus}$ (solid) and $9 \, M_{\oplus}$ (dashed) planet, relative to the case with no FeO in the mantle. These are cases where FeO is created by Si entering the core following reaction \ref{eq:Si}. The exact Fe/Mg and Si/Mg values for the low and high CMF cases are noted in the figure. }
    \label{fig:sifeo_diff}
\end{figure}

At 1 M$_\oplus$, 99.7\% of all models fall between $0.97$ and $1.05 \, R_{\oplus}$, and between $1.9$ and $2.06 \, R_{\oplus}$ at 13 M$_\oplus$ (Figure~\ref{fig:LE_wSi_MR}A).  
Using the mass as input and outputting radius, we calculate that 99.7\% of our surface gravities fall between 8.1 and $10.7 \, {\rm m} \, {\rm s}^{-2}$ at $1 \, R_{\oplus}$, and $28$ to $39 \, {\rm m} \, {\rm s}^{-2}$ at radius $\approx 2 \, R_{\oplus}$ (Figure~\ref{fig:LE_wSi_MR}B). 
We find bulk planet density varies between $\sim4.5$ and $5.7 \, {\rm g} \, {\rm cm}^{-3}$ at radius $1 \, R_{\oplus}$, and between $\approx 7.8$ and $11 \, {\rm g} \, {\rm cm}^{-3}$ at radius $\approx 2 \, R_{\oplus}$ (Figure~\ref{fig:LE_wSi_MR}C). 
We find that distributions of planet radius for planets containing 15wt\% mantle FeO due to Si entering the core are within 1\% to those produced for the simple model where all Fe remains in the core (0 wt\% mantle FeO) and no core light elements (Figure~\ref{fig:LE_wSi_MR}). 
This is despite these planets having lower CMFs compared to previous models. 
Unlike the FeO-only case with no Si in the core (Figure~\ref{fig:FeO_diffs}), as more mantle FeO is produced and Si enters the core, the size of the core must grow in radius to preserve the planet's bulk Si/Mg and Fe/Mg abundances. 
This fact, combined with the higher density of the mantle due to the incorporation of FeO, almost exactly balances out the density deficit in the core due to the presence of Si. 

We again calculated the variations in planet radius, surface gravity and bulk density due to varying amounts of mantle FeO and core Si content for the highest and lowest CMF planets in our sample, using \texttt{Perple\_X}-derived grids (Figure~\ref{fig:sifeo_diff}). 
For 1 M$_\oplus$ planets, we find radius increases by $<0.5\%$, for both the high- and low-CMF  cases, across the entire range of reasonable FeO mantle content (up to 50 wt\%), compared to the case with 0 wt\% FeO. 
Across this same range, surface gravity and bulk density decrease by $< 1\%$. 
For the $9 M_\oplus$ models, we find planet radius initially decreases slightly (by $< 0.1\%$) with increasing $w_{\rm FeO}$, up to about 20 wt\% mantle FeO, which lowers surface gravity and density by $< 0.2\%$ and $< 0.3\%$, respectively. 
Above 20 wt\% mantle FeO content, we find a maximum increase of radius of $< 1.3\%$ and decrease in surface gravity and density of $< 2.5\%$ and $< 3.6\%$, respectively at 50 wt\% mantle FeO (equivalent to $\sim 40$ wt\% Si in the core). 
Our estimated $+3\sigma$ bound of Figure~\ref{fig:LE_wSi_MR} are near the maximum density decrease expected for increasing mantle FeO content according to reaction \ref{eq:Si}, with the $-3\sigma$ bounds being nearly identical.

\subsection{Core light elements without FeO production}

Light elements can also enter into iron, and subsequently the core, prior to magma ocean creation as well \citep[e.g.,][]{Fisch20,SchlichtingYoung22}. 
Some also may be accreted directly this way, as Si is seen to alloy into Fe metal in enstatite chondrites \citep{WassonKallemeyen1988}, a likely source material of Earth.
In this case, there would be no coupled creation of oxidized FeO and reduced Si-Fe alloy (reaction~\ref{eq:Si}).
Light elements would enter the core without increasing the density of the mantle, assuming there is no subsequent reequilibration of the core with the magma ocean or mantle. 
To examine these effects, we modeled planet radius, surface gravity and bulk density, and following the same methodology as above, for planets where the density of the core is reduced by 20\%, to 80\% of that of pure Fe. 

As described in \S~\ref{sec:core}, we achieve a density reduction in the core by lowering its molar weight in accordance with the amount of light element present.
It is simple to show that as the mass fraction of the light element in the core, $w_{\rm LE}$, increases, CMF must increase according to ${\rm CMF} = \left[ 1 + (1 - w_{\rm LE}) \, \bar{\mu} / \mu_{\rm Fe} / ({\rm Fe}/{\rm Mg}) \right]^{-1}$, assuming there are no coupled reactions that will alter $\bar{\mu}$ (e.g., reaction \ref{eq:Si}). As the molar weight of the light element does not appear, any combination of light elements with the same total mass fraction would yield the same results, part of why it is difficult to distinguish which light elements are in the Earth's core. For these models, we chose O as the light element to reduce the density. 
The molar weight of a core with 10.0 wt\% O would have a mean molar weight 80\% that of pure Fe. 

\begin{figure}
    \centering
    \includegraphics[width=0.75\linewidth]{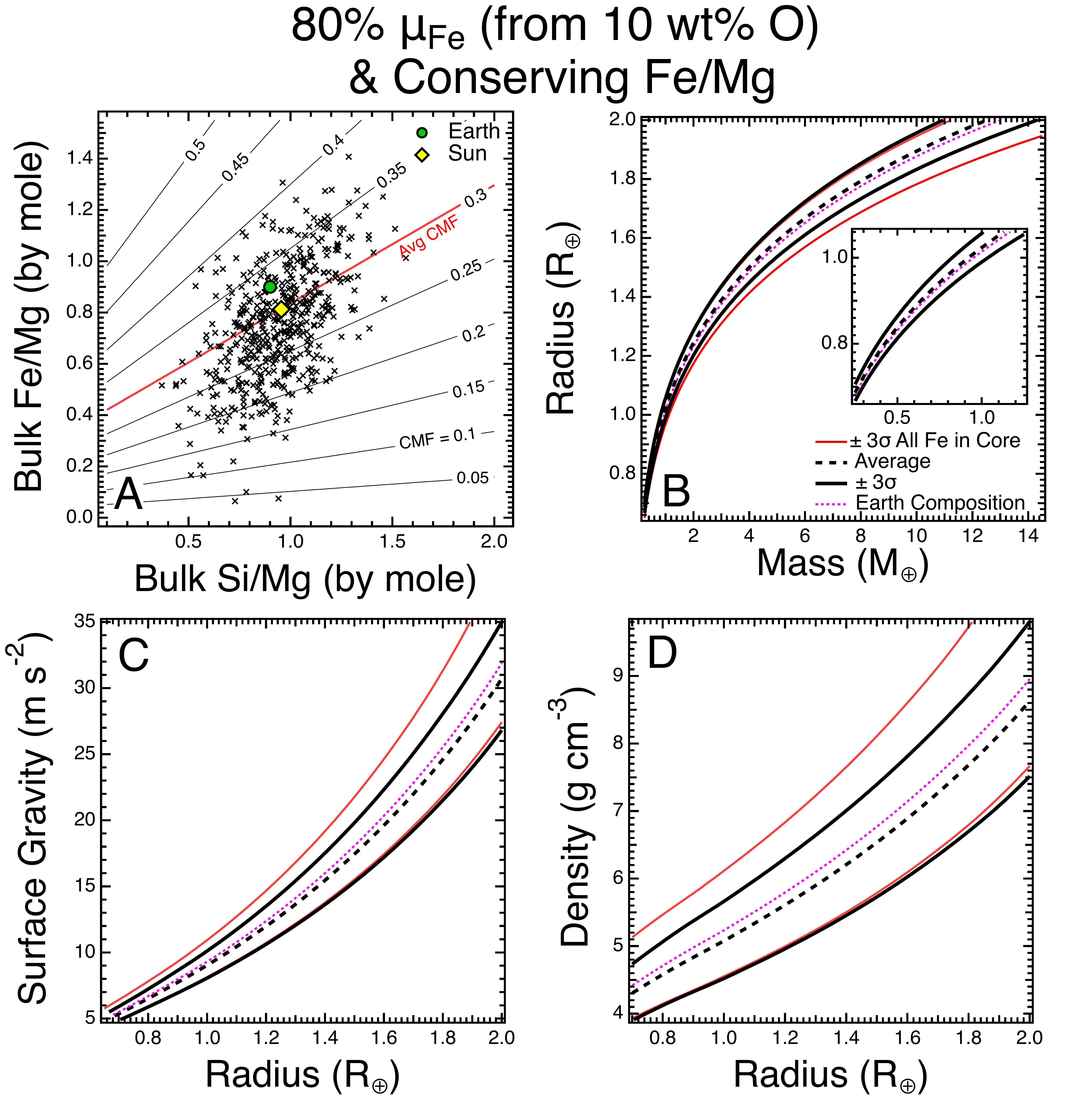}
    \caption{
    \textbf{A}: Contours of exoplanet core mass fraction (CMF) as a function of bulk Fe/Mg and Si/Mg for planets containing 10 wt\% O within the core. Markers represent the 500 abundance pairs randomly sampled from the abundance values in Figure \ref{fig:abund_hist}. Red curve is the average CMF for this distribution. \textbf{B, C, D}: Planet radius as a function of input mass (B), and surface gravity (C), and bulk density (D), as a function of planet radius (black lines), for each composition from A. In each subfigure, the output at a given input mass is fit as a Gaussian, and curves denoting the mean (dashed black) and 3$\sigma$ (99.7\%; solid black) bounds are plotted, along with a curve for an Earth-like composition (fuchsia dashed). The $\pm3\sigma$ values from the baseline model with 0 wt\% FeO and  no core light elements from Figure~\ref{fig:baseline} are shown as red lines.}
   
    \label{fig:lightel_MR}
\end{figure}

Randomly sampling 500 Fe/Mg and Si/Mg pairs from Figure \ref{fig:abund_hist}, we calculate an average CMF of 0.31 (Figure~\ref{fig:lightel_MR}), similar to our baseline and FeO-only models.
This is exactly what would be expected by replacing $\mu_{\rm Fe}$ with $0.8 \times 55.8$ g/mol, and Fe/Mg = 0.71 with $(1.11/0.8) \times 0.71$ in ${\rm CMF} = [ 1 + (\bar{\mu}/\mu_{\rm Fe}) / ({\rm Fe}/{\rm Mg})]^{-1}$, to produce a core that has the same total number of Fe atoms, but with 10.0 wt\% O. 
At 1 M$_\oplus$, 99.7\% of all models fall between $0.99$ and $1.06 \, R_{\oplus}$, and between $1.95$ and $2.08 \, R_{\oplus}$ at 13 M$_\oplus$ (Figure~\ref{fig:lightel_MR}B).  
Using the mass as input and outputting radius, we calculate that 99.7\% of our surface gravities fall between 8.1 and $10 \, {\rm m} \, {\rm s}^{-2}$ at $1 \, R_{\oplus}$, and $27$ to $35 \, {\rm m} \, {\rm s}^{-2}$ at radius $\approx 2 \, R_{\oplus}$ (Figure~\ref{fig:lightel_MR}C). 
We find bulk planet density varies between $\sim4.5$ and $5.6 \, {\rm g} \, {\rm cm}^{-3}$ at radius $1 \, R_{\oplus}$, and between $\approx 7.5$ and $9.8 \, {\rm g} \, {\rm cm}^{-3}$ at radius $\approx 2 \, R_{\oplus}$ (Figure~\ref{fig:lightel_MR}D). 
We find that, in general, planets with 10.0 wt\% O in the core (but without FeO in the mantle) have a larger radius for the same mass, and smaller surface gravity and bulk density, compared to the cases of pure Fe cores
(Figure~\ref{fig:lightel_MR}). 
Our estimated -3$\sigma$ bounds of each quantity are nearly identical to those of the baseline case from Figure~\ref{fig:baseline} differing by $<1$\%. 
The $+3\sigma$ bound for planet radius across these compositions is $\leq3\%$ greater than the baseline case up to 14.5 M$_\oplus$. 
This equates to maximum decreases in surface gravity and density of $\sim$6\% and 8\% decreases in surface gravity and bulk density, respectively. 

\begin{figure}
    \centering
    \includegraphics[width=0.75\linewidth]{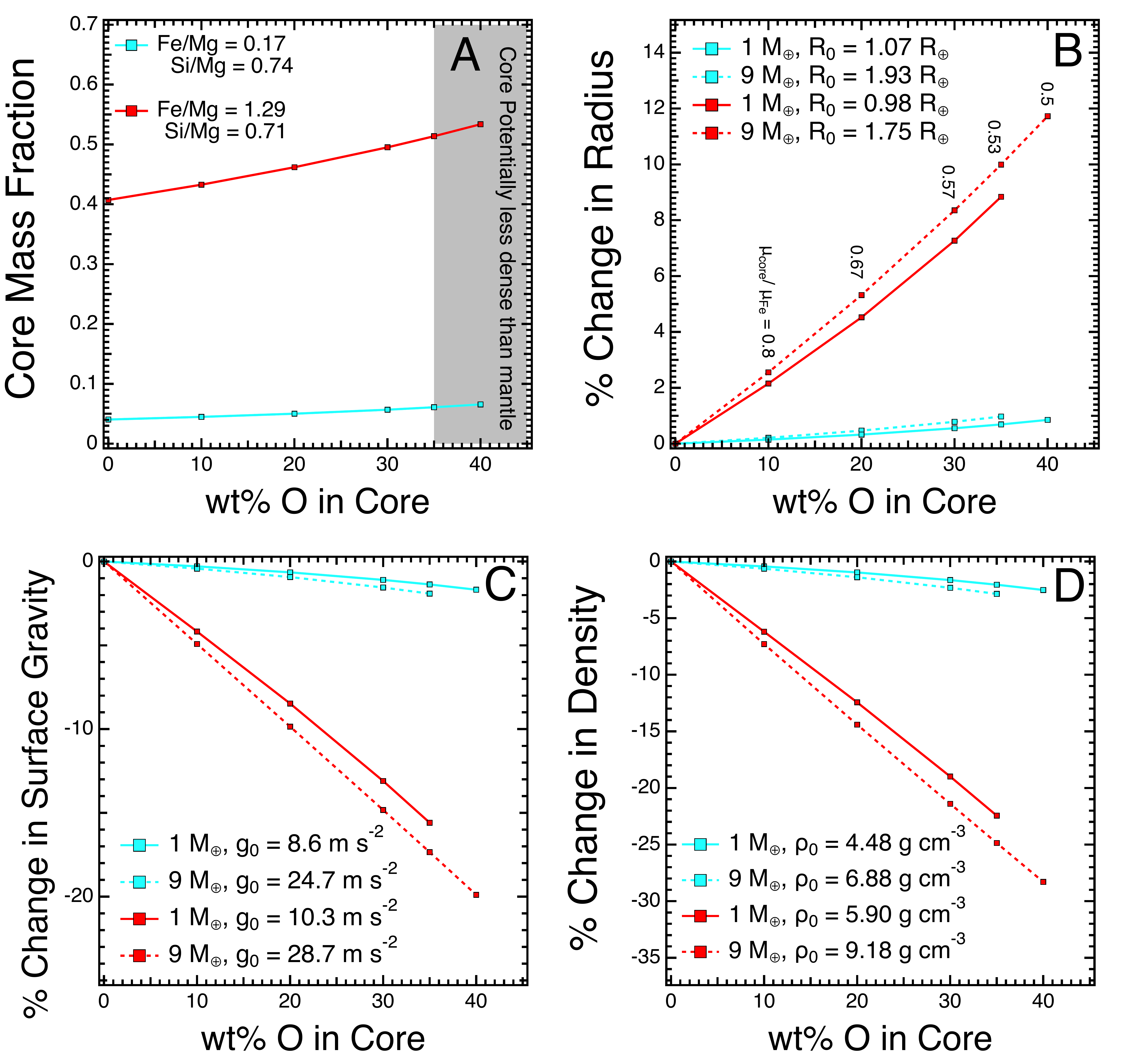}
    \caption{Core mass fraction (\textbf{A)}, percent difference in planet radius (\textbf{B}), surface gravity (\textbf{C}) and density (\textbf{D}), as functions of O mass fraction in the core, for our lowest-CMF (teal) and highest-CMF (red) models in a $1 \, M_{\oplus}$ (solid) and $9 \, M_{\oplus}$ (dashed) planet, relative to the case with no O in the core. These cases assume no FeO in the mantle. The exact Fe/Mg and Si/Mg values for the low and high CMF cases are noted in the figure. }
    \label{fig:O_core}
\end{figure}

Treating the mass fraction of O in the core as a free parameter, we find that the core density becomes less than the mantle density if this fraction exceeds $\approx 40{\rm wt}\%$ O (Figure~\ref{fig:O_core}). 
Below 40 wt\% O, we find planet radius increases as the mass fraction of O in the core increases, reaching maximum increases of $\approx 2\%$ and $13\%$ for the bulk Fe-poor and Fe-rich cases, respectively, compared to the case with no O in the core. 
The corresponding decrease in surface gravity is $\sim2$\% and 20\% for the Fe-poor and Fe-rich cases, respectively.
Bulk density likewise decreases by a maximum of $\sim$3\% and 28\% for the Fe-poor and Fe-rich cases, respectively. 
Our -3$\sigma$ bound in planetary radius from Figure \ref{fig:lightel_MR}, therefore, is nearly the same as the maximum decrease expected for physically-likely compositions. 
The +3$\sigma$ bound in radius, would increase as the mass fraction of light elements within the core increases, while still remaining within the range of radii outlined by our baseline case (red lines, Figure \ref{fig:lightel_MR})

\section{Effects of volatile layers on planetary mass and radius}
\label{sec:water}
Volatile layers (e.g., liquid water, or H$_2$ gas) on the surface of a rocky exoplanet also lower the density of an exoplanet compared to the volatile-free case \citep[e.g.,][]{Seag07, Gill16, Gill17, Unter18,haldemann_aqua_2020}. 
To quantify how the presence of volatile layers affects the density of planet, we again randomly sample 500 Fe/Mg and Si/Mg compositions as inputs into the ExoPlex code and model across masses up to $\approx \, 12 \, M_\oplus$ (roughly that of a $2 \, R_\oplus$ planet). 
For simplicity, we simply assume the planet has all Fe within its core, no mantle FeO and no core light elements. 
As expected, these compositions yield a range of CMF---as defined by the mass of the core relative to only the core and silicate layers---from 0.10 to 0.46, similar to those in Figure~ \ref{fig:baseline}. 

\begin{figure}
    \centering
    \includegraphics[width=0.75\linewidth]{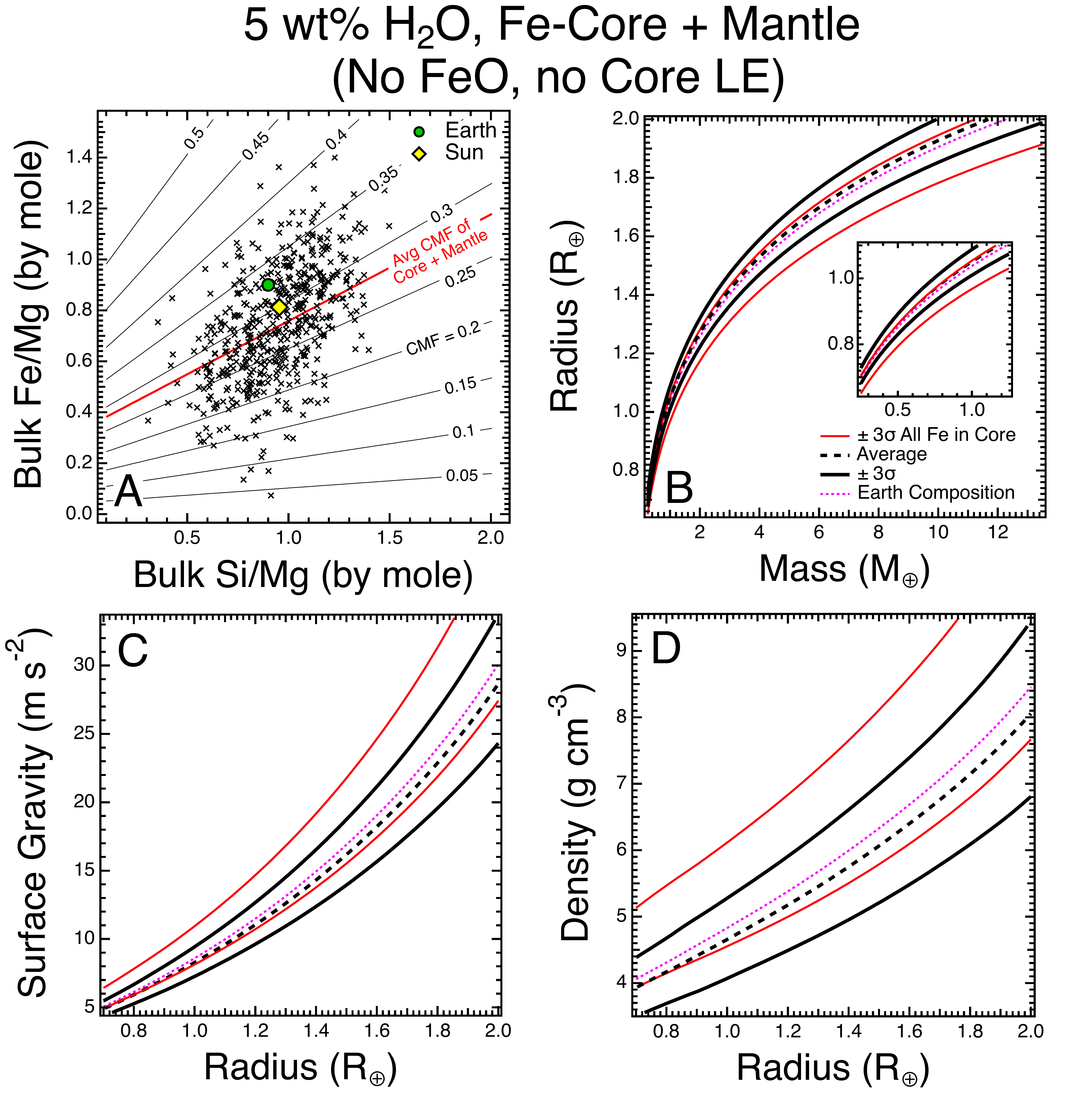}
    
    \caption{
        \textbf{A}: Contours of exoplanet core mass fraction (CMF) as a function of bulk Fe/Mg and Si/Mg for planets with 5 wt\% surface water assuming all Fe remains in the core. Markers represent the 500 abundance pairs randomly sampled from the abundance values in Figure \ref{fig:abund_hist}. Red curve is the average CMF for this distribution. \textbf{B, C, D}: Planet radius as a function of input mass (B), and surface gravity (C), and bulk density (D), as a function of planet radius (black lines), for each composition from A. In each subfigure, the output at a given input mass is fit as a Gaussian, and curves denoting the mean (dashed black) and 3$\sigma$ (99.7\%; solid black) bounds are plotted, along with a curve for an Earth-like composition (fuchsia dashed). The $\pm3\sigma$ values from the baseline model with 0 wt\% FeO and  no core light elements from Figure~\ref{fig:baseline} are shown as red lines.}    
    \label{fig:water}
\end{figure}

We choose water/ice as our surface volatile, and set its abundance to 5 wt\% of the planet.
We continue to set the potential temperature in the mantle to 1600 K, and set the potential temperature in the water layer to 300 K.
At 1 M$_\oplus$, 99.7\% of all models fall between $1.01$ and $1.09 \, R_{\oplus}$, and between $1.97$ and $2.13 \, R_{\oplus}$ at 13 M$_\oplus$ (Figure~\ref{fig:water}B).  
Using the mass as input and outputting radius, we calculate that 99.7\% of our surface gravities fall between 7.3 and $9.6 \, {\rm m} \, {\rm s}^{-2}$ at $1 \, R_{\oplus}$, and $24$ to $33 \, {\rm m} \, {\rm s}^{-2}$ at radius $\approx 2 \, R_{\oplus}$ (Figure~\ref{fig:water}C). 
We find bulk planet density varies between $\sim4.1$ and $5.3 \, {\rm g} \, {\rm cm}^{-3}$ at radius $1 \, R_{\oplus}$, and between $\approx 6.8$ and $9.4 \, {\rm g} \, {\rm cm}^{-3}$ at radius $\approx 2 \, R_{\oplus}$ (Figure~\ref{fig:water}D). 
The addition of 5 wt\% water, therefore, increases the radius of the planet by $\sim 0.05--0.1  \, R_\oplus$, which decreases a planet's surface gravity and density. 
Unlike previous models, the bulk density of the planets with the lowest CMF lie below the $-3\sigma$ bounds of the baseline model and below that of a pure rock sphere. 
This is because the addition of this volatile layer increases the radius of the planet while adding only minimal mass, due to its much lower density ($\sim 1 \, {\rm g} \, {\rm cm}^{-3}$) compared to rock ($> 3 \, {\rm g} \, {\rm cm}^{-3}$) and iron ($> 8 \, {\rm g} \, {\rm cm}^{-3}$).  

\begin{figure}
    \centering
    \includegraphics[width=\linewidth]{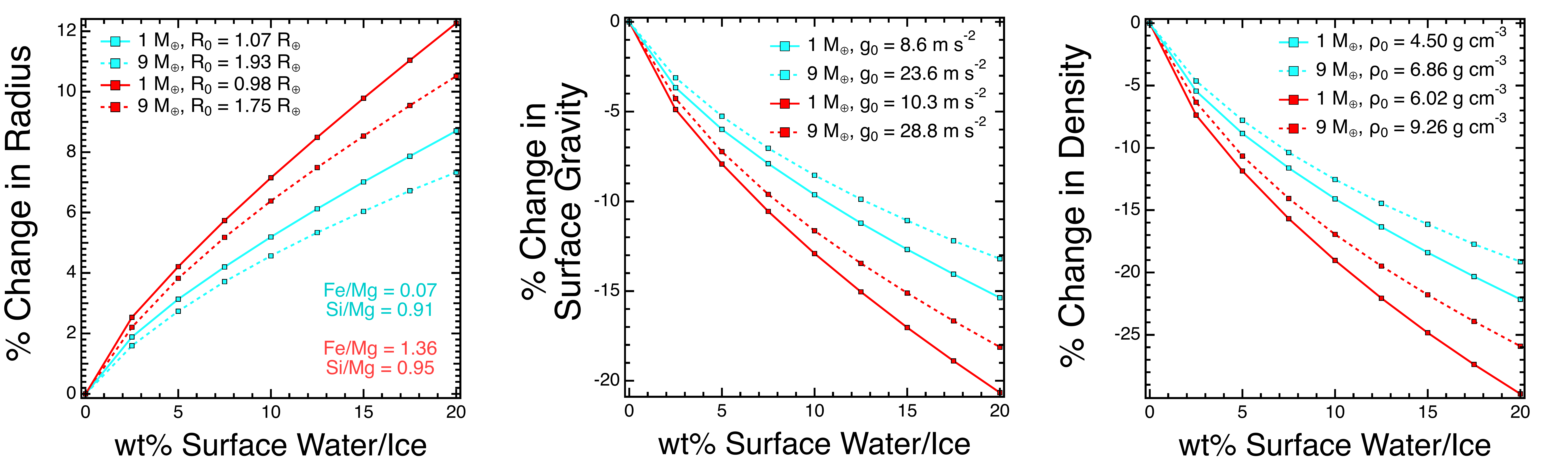}
    \caption{Fractional change in planet radius (left), surface gravity (center) and density (right) with water ice content, for our lowest-CMF (black) and highest-CMF (red) models in $1 \, M_{\oplus}$ (solid) and $9 \, M_{\oplus}$ (dashed) planets, relative to the case with no water/ice. The exact Fe/Mg and Si/Mg values for the low and high CMF cases are noted in the figure. }
    \label{fig:water_diff}
\end{figure}

Examining the effects of water contents on the highest and lowest CMF compositions in our sample, we find that as the mass fraction of water on a planet increases, planet radius increases; the increase is as much as 12\% for a planet with 20 wt\% surface water (Figure~\ref{fig:water_diff}).
At higher planet masses, however, this increase in radius is muted slightly due to the increased pressure in the water/ice layers, which increases their density according to the compressibility of water ice. These larger radii subsequently lower the surface gravity and bulk density of these planets, by as much as $\approx 21\%$ and 30\%, in the 1 M$_\oplus$, high-CMF case. 
We estimate that for $1 \, M_{\oplus}$ and $9 \, M_{\oplus}$ planets, when water contents reach $\approx 12$ wt\% and $\approx 17$ wt\%, respectively, each of the 500 randomly-sampled Fe/Mg and Si/Mg compositions will be completely below outside the 3$\sigma$ bounds of radius, surface gravity and bulk density of our baseline pure Fe-core model (red bands, Figure~\ref{fig:water}). 

\section{Degeneracy in mass-radius-composition models}
\label{sec:degen}
\begin{figure}
    \centering
    \includegraphics[width=.9\linewidth]{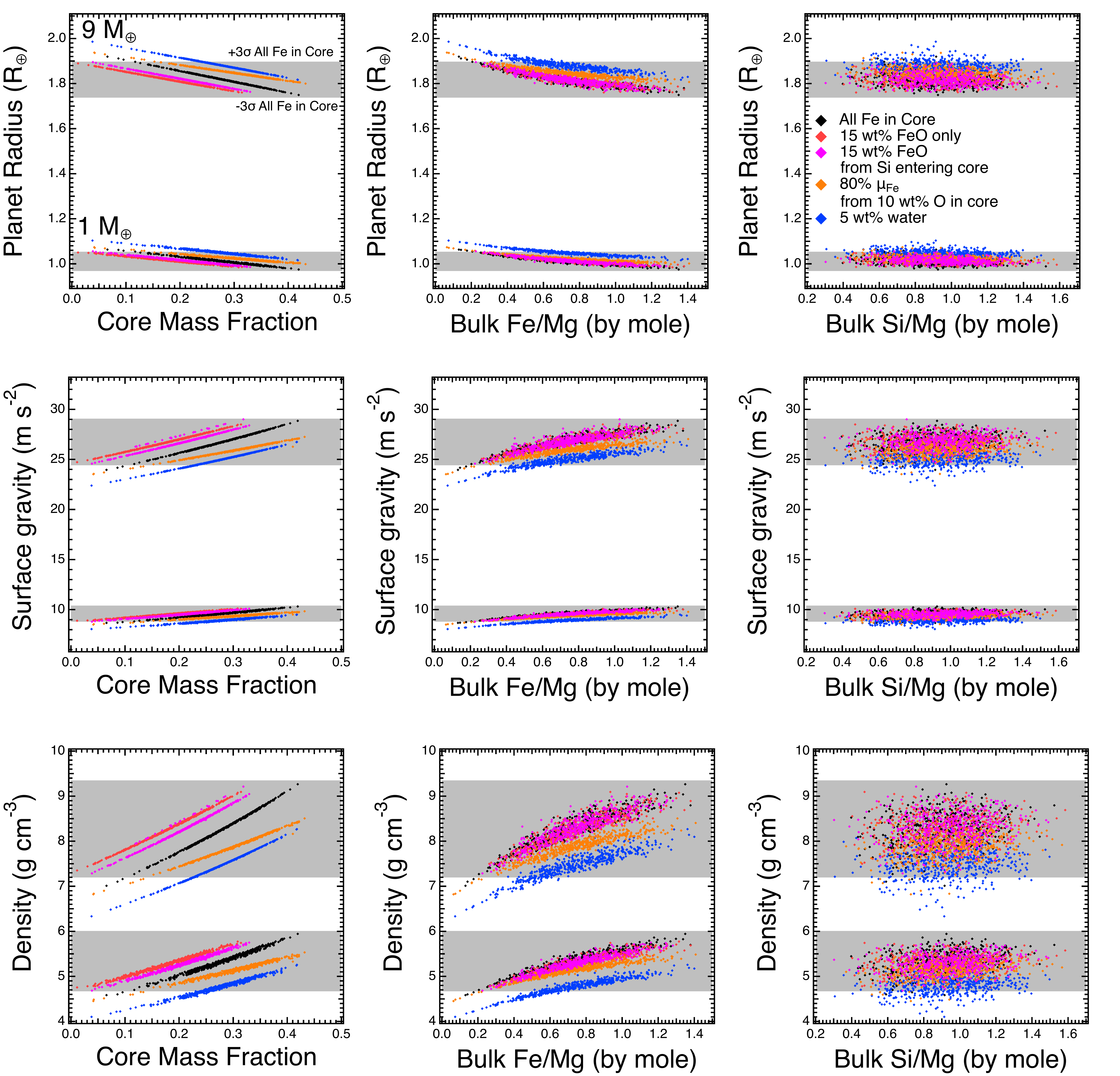}
    \caption{Planetary radius (top), surface gravity (center) and bulk density (bottom) as a function of core mass fraction (left), bulk Fe/Mg (center) and bulk Si/Mg (right). Colors represent the different compositional cases explored in this work (as in Figure 14; see top right panel for key). Gray bands represent the $\pm3\sigma$ confidence bands for the baseline model with all Fe in the core.}
    \label{fig:CMF_compare}
\end{figure}

Characterizing an individual exoplanet's composition is difficult due to the underlying uncertainty in mass and radius measurements, but also the inherent degeneracy in mass-radius models, with multiple compositions producing planets of similar mass or radius \citep[e.g., ][]{Valencia2007, Rogers_2010, Dorn15, Unter16}. 
Figure~\ref{fig:CMF_compare} shows this degeneracy more broadly, where the resulting range of radii, surface gravities and bulk densities for 1 and 9 M$_\oplus$ planets for each of our model compositions overlap considerably, despite these planets having very different oxidation states, core mass fractions and silicate mantle sizes.

We find planetary radius, surface gravity and density are linearly proportional to both a planet's CMF and mass, however, multiple CMF values produce planets of the same radius for a given mass (Figure~\ref{fig:CMF_compare}, left column). 
For example, a $1 \, M_\oplus$, $1 \, R_\oplus$ planet could have CMF anywhere between $\approx 0.21$, if it contains 15 wt\% FeO in its mantle, to $\approx 0.42$, if its core is 10wt\% O, a considerable difference. 
These derived values are also each linearly proportional to the planet's bulk Fe/Mg abundance, albeit slightly less coherently than CMF (Figure~\ref{fig:CMF_compare}, center column). 
There is, however, less degeneracy in the radius, bulk Fe/Mg and mass parameter space than CMF across our different compositional cases. 
For a $1 \, M_\oplus$ planet, the calculated radii for each of the compositional cases (besides the 5wt\% water case) nearly overlap for a given bulk Fe/Mg. 
We find similar results at $9 \, M_\oplus$.
However, the model with 10 wt\% O in the core case does produce slightly higher radii than the other non-water models for a given Fe/Mg. 
None of our models are particularly sensitive to the planet's bulk Si/Mg (Figure~\ref{fig:CMF_compare}, right column), in agreement with previous work \citep[e.g., ][]{Dorn15, Unter16}. 
CMF or bulk Si/Mg, therefore, then, are unlikely to be inferred from mass and radius measurements alone and are thus  are not adequate to describe a planet's composition without additional compositional constraints. 
A planet's bulk Fe/Mg, however, might potentially be observable if the presence of core light elements can be ruled out.   

\begin{figure}
    \centering
    \includegraphics[width=\linewidth]{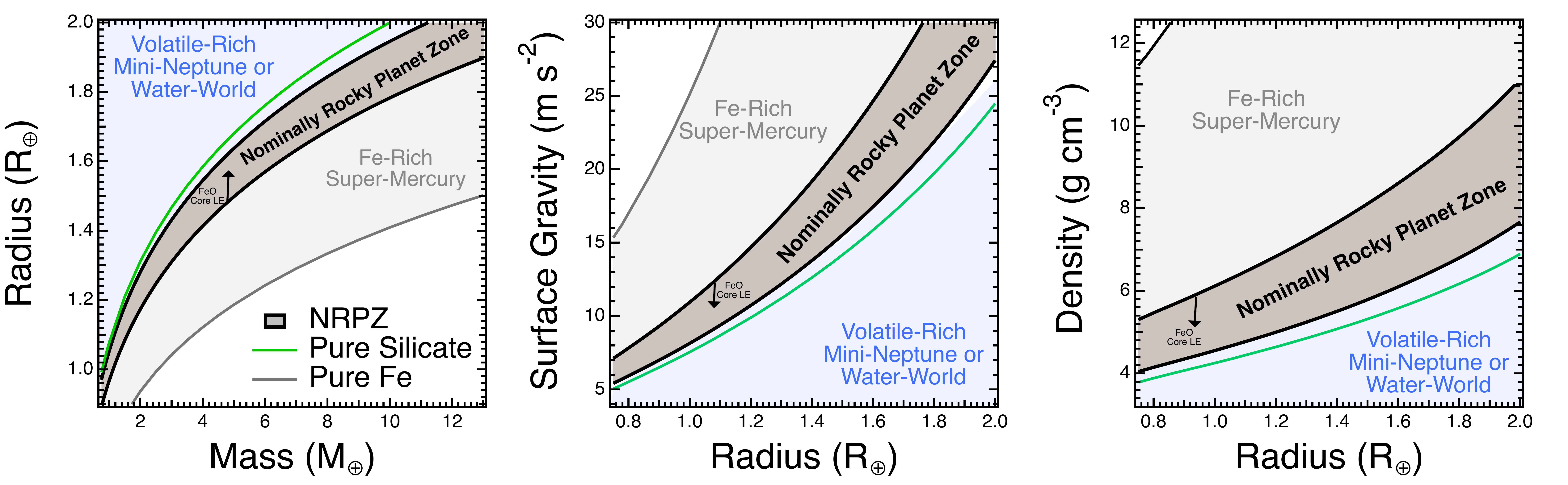}
    \caption{The $\pm3\sigma$ ranges of planet radius as a function of mass (left), and surface gravity (center) and bulk density (right) as functions of planet radius, of the nominally rocky planet zone (NRPZ, black curves). We define the NPRZ as our  baseline case with all Fe in the core with no light elements and no mantle FeO. All other compositional cases generally increase a planet's radius, and decrease its surface gravity and density, for a given mass (Figures \ref{fig:feo_per}, \ref{fig:sifeo_diff} and \ref{fig:O_core}). The cases of a pure Fe planet (thin black curve) and a pure silicate planet (green curve) are included for reference.}
    \label{fig:degen}
\end{figure}


\section{The Nominally Rocky Planet Zone}
\label{sec:nrpz}
Unfortunately, this degeneracy limits our ability to resolve the non-Fe elemental compositions and mineralogy of all but the best resolved exoplanets. 
Our best-fit $\pm3\sigma$ bounds in radius, surface gravity and bulk density of each of our water-free compositional cases are nearly identical or wholly within the baseline case where all Fe resides in the core (Figures \ref{fig:FeO_diffs}, \ref{fig:LE_wSi_MR}, \ref{fig:lightel_MR}).
Even when the respective chemical components in each model are treated as free parameters, the $+3\sigma$ bounds in radius vary less than $\sim$1\% at extreme mantle FeO contents or core light element budgets (Figures \ref{fig:feo_per}, \ref{fig:sifeo_diff}, \ref{fig:O_core}). 
The $-3\sigma$ bounds in radius also increases as these chemical parameters increase, however this only narrows the range of predicted radii for planets of a given mass relative to the baseline case.
This means the distribution of radii predicted in the simple chemical case of all Fe remaining within the core and without mantle FeO or core light elements (Figure \ref{fig:baseline}) represents the widest range of radii a rocky planet without significant surface volatiles. 
Each of these models were created under the well-grounded assumption that the distribution of all available stellar abundances of the primary rocky planet-building elements represents the distribution of these abundances in the interiors of rocky exoplanets (see \S\ref{sec:stellar}). 
We therefore define those planets that fall within the range of radii, surface gravity and bulk density of our baseline case to be ``nominally rocky,'' meaning their observed mass and radius alone are consistent with the planet having a bulk composition consistent with the distribution of stellar abundance data without the addition of significant surface volatiles.
Planets with measured densities outside of this ``nominally rocky planet zone'' (NRPZ; Figure \ref{fig:degen}) would therefore require either an anomalous bulk composition relative to the stellar abundance data or the addition of other compositional layers (e.g., melt, volatiles). 
The NRPZ, therefore, also provides us with a quantitative tool to identify those planets that are definitively \textit{not} rocky without the need for host-star abundance data. 

Examining the lower bound in radius (upper bound for gravity and density) of tne NRPZ first, these planets would require significant enrichment in Fe relative to the range of stellar compositions to account for their increased density. 
These planets are the so-called ``super-Mercury'' exoplanets. 
We assert then, that any planets with surface gravity ($g$) or density $\rho$ that are above the upper bound in radius and lower bounds in surface gravity and density of the NRPZ (Figure~\ref{fig:degen}), are very likely then super-Mercuries .
Fitting this curve to planets between with radii $R$ between 0.8 and $2 \, R_{\oplus}$, we can infer the existence of a super-Mercury (SM) if its surface gravity $g$, density $\rho$ or mass $M$ exceed the maximum possible for a nominally rocky exoplanet, i.e., if 
\begin{equation}
    g > g_{\rm SM} = \left[ 4.3 + 6.6\left(\frac{R}{R_\oplus}\right)^{2.4} \right] \, \mathrm{m\ s^{-2}},
\end{equation}
and 
\begin{equation}
    \rho > \rho_{\rm SM} = \left[ 4.6 + 1.5\left(\frac{R}{R_\oplus}\right)^{2.1} \right] \, \mathrm{g\ cm^{-3}}.
\end{equation}
and 
\begin{equation}
  	M > M_{SM} = \left[ 0.25 + 0.87\left(\frac{R}{R_\oplus}\right)^{4.2} \right] \, \mathrm{M_{\oplus}}.
\end{equation}
Exoplanets exceeding these limits must have Fe concentrations well above the typical range of Fe/Mg values in stars in Figure~\ref{fig:abund_hist}, either because their host star is truly anomalous, or because the planet became enriched in Fe relative to Mg due to a mantle stripping during an impact
\citep[e.g.,][]{Bonomo19} or other unknown processes. 

Examining the NRPZ's upper bound of radius (lower bound of surface gravity and bulk density), a planet with larger radius must be enriched in low-density materials such as surface volatiles, to explain its observed mass and radius. 
It likely then to be a mini-Neptune (if enriched in an H$_2$ atmosphere) or water world (if enriched in H$_2$O).
Fitting this boundary to planets between with radii $R$ between 0.8 and $2 \, R_{\oplus}$, one could infer the existence of a volatile-rich mini-Neptune/water world (MN-WW) if its surface gravity $g$, density $\rho$ or mass $M$ lie below the minimum values possible for a nominally rocky exoplanet, i.e., if 
\begin{equation}
    g < g_{\rm MN-WW} = \left[ 3.3 + 4.7\left(\frac{R}{R_\oplus}\right)^{2.3} \right] \, \mathrm{m\ s^{-2}}
\end{equation}
or
\begin{equation}
    \rho < \rho_{\rm MN-WW} = \left[ 3.6 + 0.91\left(\frac{R}{R_\oplus}\right)^{2.1} \right] \,  \mathrm{g\ cm^{-3}}.
\end{equation}
or if 
\begin{equation}
  	M < M_{MN-WW} = \left[0.19 + 0.64\left(\frac{R}{R_\oplus}\right)^{4.1} \right] \, \mathrm{M_{\oplus}}.
\end{equation}
A planet satisfying these criteria could be a rocky exoplanet if its host star were anomalously Fe-poor, but is more likely to be volatile-rich \citep[e.g., ][]{Brink22}.
We note that the presence of hydrated molten silicate mantle will lower may also increase a planet's radius as well \citep{Dorn21}.
A planet with an anhydrous silicate magma ocean is not likely to have a radius much larger than if it were solid, however, due to the fact that the density of silicate melt approaches that of the solid at moderate pressures in the magma ocean \citep{Cara19}.
Planets with measured radii above NRPZ are still likely volatile-rich, although perhaps more exotic than a mini-Neptune or water-world. 
We note those planets outside of the NRPZ at the $\geq1\sigma$ level that have equilibrium temperatures above the zero-pressure melting curve of dry peridotite \citep{Katz2003} in Table \ref{tab:NPRZ}.

\begin{figure}
    \centering
    \includegraphics[width=.5\linewidth]{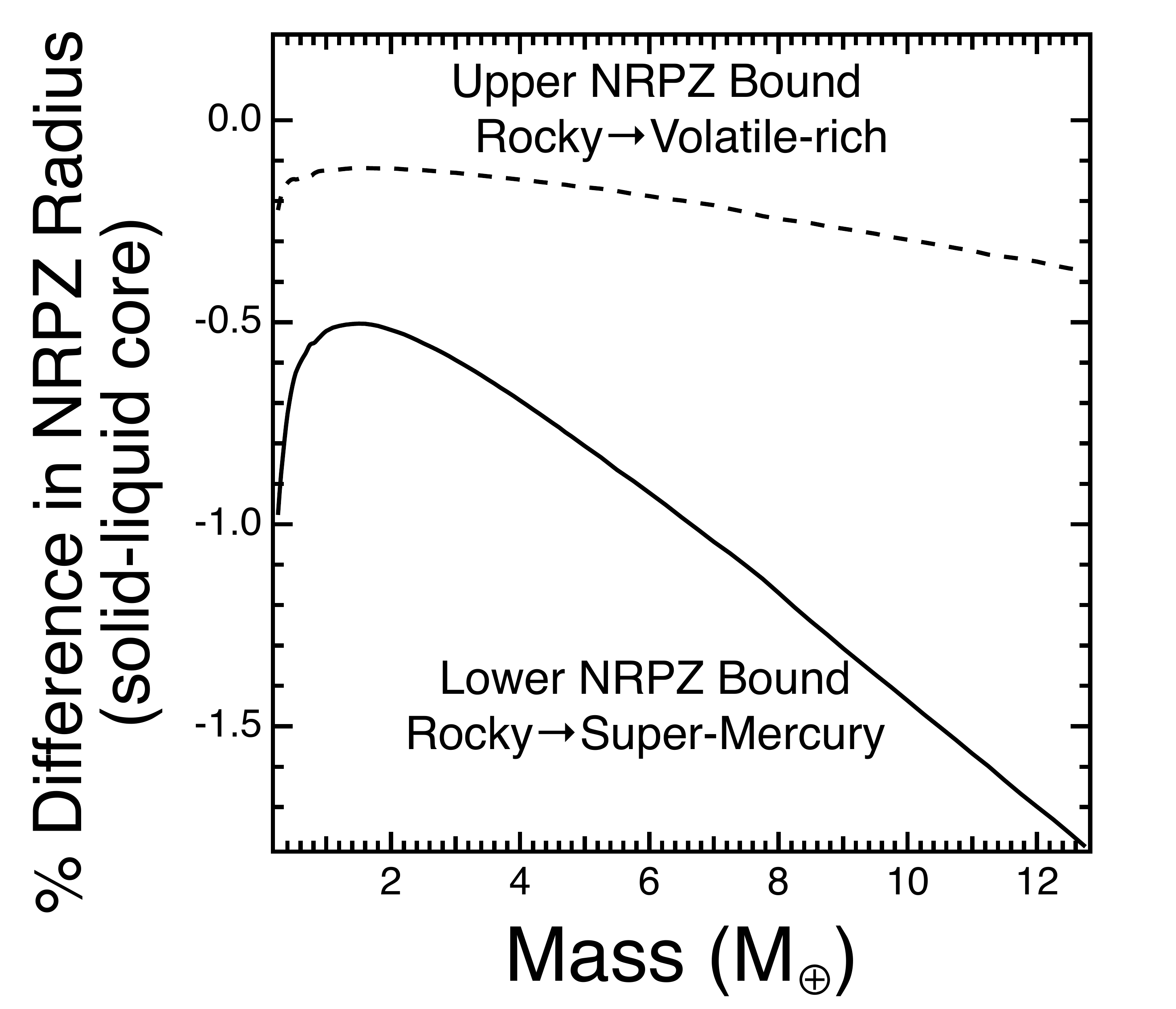}
    \caption{Percent difference between for models assuming an entirely liquid and solid central Fe-core for the lower- (solid) and upper- bounds of the NRPZ.}
    \label{fig:solid_fe}
\end{figure}

Our model of the NRPZ assumes the central Fe-core is entirely liquid. 
Solid iron, however, is potentially stable at the pressure and temperatures within rocky exoplanets \citep[e.g., ][]{Bouj20}.
We examine the effects on the location of the NRPZ by rerunning our baseline planet model assuming an entirely solid Fe-core where all Fe resides in the core for the same 500 Fe/Mg and Si/Mg compositions from figure \ref{fig:baseline}. 
We then again fit the 99.7\% ($3\sigma)$ bounds of the resulting radius distribution for a given input mass by fitting to a Gaussian.
For these models we adopt the Vinet equation of state for $\epsilon$-Fe of \citet{Dewaele17} for planet masses up to 13 M$_\oplus$ and calculate density directly using BurnMan.
Above 13 M$_\oplus$, the pressure within a planet's mantle begins to exceed our maximum grid pressure of 2.8 TPa, particularly for those planets with CMF lower than Earth. 
We find the upper- and lower-limits of the NRPZ for radius decrease by $<0.4$ and $<1.8\%$, respectively (Figure \ref{fig:solid_fe}). 
These values are likely upper-limits, however, as planetary core are likely a mixture of solid and liquid Fe \citep{Bouj20}. 
Additionally, the EoS of \citet{Dewaele17} is isothermal with a temperature of 300 K. 
Higher temperatures will lower the density of $\epsilon$-Fe, further reducing the change in the radius values of the NRPZ compared to our liquid Fe model.
Uncertainties within the underlying equations of state of the core or mantle are unlikely to affect the location of the NRPZ as they have small effect on a planet's bulk density \citep{Unter19}.

\begin{figure}[t!]
    \centering
    \includegraphics[width=\linewidth]{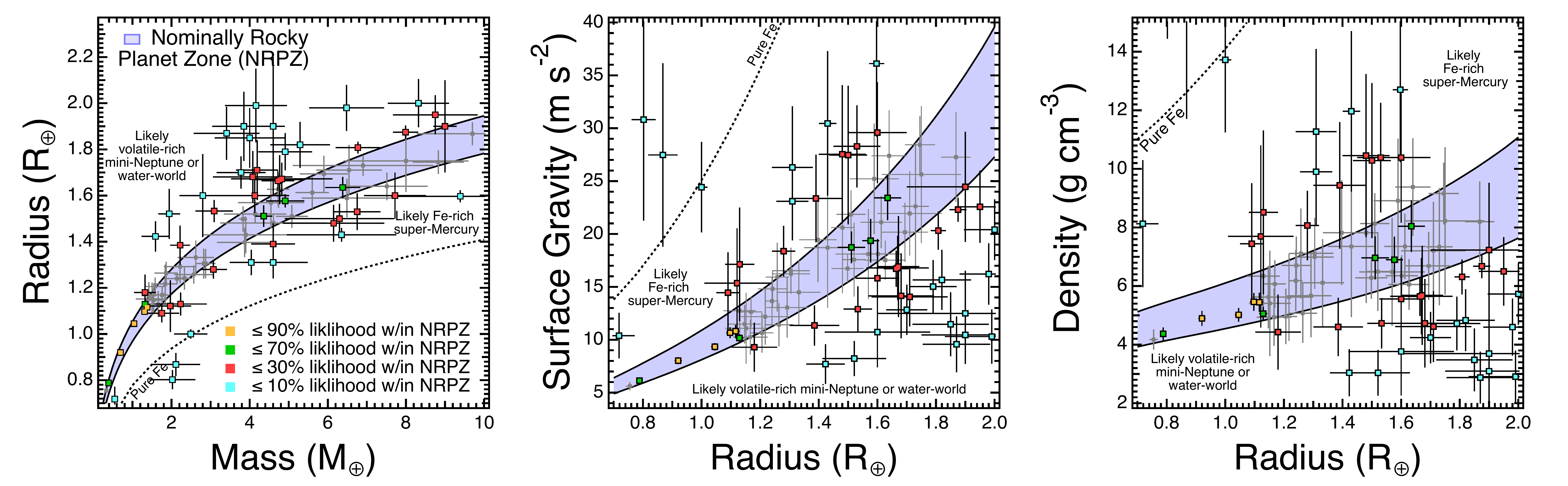}
    \caption{Planet mass and radius (left), surface gravity (center) and bulk density (right) for 85 exoplanets with uncertainties in mass and radius $<30\%$ (gray). The nominally rocky planet zone (NRPZ) is shaded blue. Those planets with likelihoods of falling within the NRPZ at $\geq90\%$ and 70--90\% confidence are marked orange and green, respectively. Those planets with likelihoods of falling \textit{outside} of the NRPZ at $\geq90\%$ and 70--90\% confidence are marked teal and red, respectively.}
    \label{fig:planets}
\end{figure}

Few small ($R < 2 \, R_{\oplus}$) exoplanets can be definitively identified as rocky planets like Earth.
There are 86 small exoplanets with measured uncertainties in both radius and mass less than 30\% within the NASA Exoplanet Archive as of time of publication\footnote{accessed 11/2/2022} \citep[][]{Akeson13},\dataset[10.26133/NEA1]{https://doi.org/10.26133/NEA1}) .
To assess the probability these exoplanets are within the NRPZ, we sampled $10^5$ mass-radius pairs within their respective error ellipses and determined the fraction of these pairs that fell within the NRPZ.
We find only four planets have masses and radii and low respective uncertainties, consistent with being within the NRPZ at the 90\% confidence level (TRAPPIST-1 b, c, e and f); and five more (TRAPPIST-1 d and g, GJ 9827 b, HD 219134 c and LHS 1140 b) consistent between 70--90\% confidence (Figure~\ref{fig:planets}, Table~\ref{tab:NPRZ}). 
Each of these planets have uncertainties in radius and mass of $< 5\%$ and $< 10\%$, respectively, lending to their high probability of lying within the NRPZ.

More planets actually can be ruled out from being rocky exoplanets.
We estimate 20 planets are consistent with being outside of the NRPZ at $\geq$90\% confidence (these planets have <10\% likelihood of being within the NRPZ) and and additional 23 at between 70--90\% confidence (Figure~\ref{fig:planets}, Table~\ref{tab:NPRZ}).
These planets are less well-resolved, with uncertainties up to 16\% in radius and 26\% in mass. 
To definitively classify an exoplanet as rocky requires very small uncertainties in mass and radius, but definitively classifying them as being \textit{not} rocky, either as a super-Mercury, volatile-rich world does not. 
Of these 43 non-rocky exoplanets, we estimate 17 are likely Fe-enriched super-Mercuries, and 26 likely are volatile-enriched and potentially a mini-Neptune/water-world (Table \ref{tab:NPRZ}). 

Defining the bounds of the NRPZ in this manner, using stellar abundance data, therefore allows us to classify the 85 small exoplanets as follows: 9 are consistent within the NRPZ; 43 are definitely {\it not} rocky with Earth-like CMF (17 are super-Mercuries and 26 are volatile-enriched); and 36 cannot yet be classified.
The majority of the unclassified planets lie near the boundaries of the NRPZ, where smaller uncertainties are required to distinguish whether a planet is within or outside the NRPZ.

\begin{deluxetable}{lccccc}
\tablecolumns{6}
\vspace{-30pt}
\tablewidth{0pt}
\tabletypesize{\scriptsize}
\tablecaption{Sample of planets with uncertainties less than 30\% in both mass and radius with the highest and lowest likelihood of being in NRPZ} \label{tab:NPRZ}
\tablehead{
         \colhead{} & \colhead{Radius} &\colhead{Mass}  &  \colhead{Bulk Density} & \colhead{Surface Gravity} & \colhead{Probability} \\
                  \colhead{Planet} & \colhead{(R$_{\oplus}$)} &\colhead{(M$_{\oplus}$)} & \colhead{(g cm$^{3}$)}  & \colhead{(m s$^{-2}$)}  & \colhead{w/in NRPZ} 
}
\startdata
        \multicolumn{6}{c}{$\geq70\%$ Likelihood of being within NRPZ}\\
        \hline
        TRAPPIST-1 c & 1.10 $\pm$ 0.01 & 1.31 $\pm$ 0.06 & 5.46 $\pm$ 0.30 & 10.66 $\pm$ 0.52 & 99 \\ 
        TRAPPIST-1 b & 1.12 $\pm$ 0.01 & 1.37 $\pm$ 0.07 & 5.45 $\pm$ 0.33 & 10.82 $\pm$ 0.60 & 98 \\ 
        TRAPPIST-1 f & 1.05 $\pm$ 0.01 & 1.04 $\pm$ 0.03 & 5.02 $\pm$ 0.23 & 9.33 $\pm$ 0.36 & 95 \\ 
        TRAPPIST-1 e & 0.92 $\pm$ 0.01 & 0.69 $\pm$ 0.02 & 4.90 $\pm$ 0.25 & 8.02 $\pm$ 0.34 & 92 \\ 
        TRAPPIST-1 g & 1.13 $\pm$ 0.01 & 1.32 $\pm$ 0.04 & 5.06 $\pm$ 0.24 & 10.17 $\pm$ 0.39 & 88 \\ 
        GJ 9827 b & 1.58 $\pm$ 0.03 & 4.91 $\pm$ 0.49 & 6.90 $\pm$ 0.79 & 19.37 $\pm$ 2.06 & 80 \\ 
        HD 219134 c & 1.51 $\pm$ 0.05 & 4.36 $\pm$ 0.22 & 6.97 $\pm$ 0.74 & 18.73 $\pm$ 1.50 & 80 \\ 
        TRAPPIST-1 d & 0.79 $\pm$ 0.01 & 0.39 $\pm$ 0.01 & 4.37 $\pm$ 0.22 & 6.13 $\pm$ 0.25 & 79 \\ 
        LHS 1140 b & 1.64 $\pm$ 0.05 & 6.38 $\pm$ 0.45 & 8.05 $\pm$ 0.88 & 23.41 $\pm$ 2.11 & 75 \\ 
        \hline
        \multicolumn{6}{c}{$\leq30\%$ Likelihood of being within NRPZ}\\
        \hline
        Kepler-80 e & 1.60$\pm$0.08 & 4.13$\pm$0.81 & 5.56$\pm$1.37 & 15.83$\pm$3.48 & 30$^{\dagger}$ \\ 
        KOI-1599.02 & 1.90$\pm$0.20 & 9.00$\pm$0.30 & 7.23$\pm$2.30 & 24.46$\pm$5.21 & 29$^{\dagger}$ \\ 
        TOI-1452 b & 1.67 $\pm$0.07 & 4.82 $\pm$1.30 & 5.68 $\pm$1.70 & 16.91 $\pm$4.78 & 28$^\dagger$ \\ 
        GJ 1252 b & 1.18$\pm$0.08 & 1.32$\pm$0.28 & 4.43$\pm$1.29 & 9.30$\pm$2.32 & 28$^{\dagger}$ \\ 
        TOI-178 c & 1.67$\pm$0.11 & 4.77$\pm$0.55 & 5.65$\pm$1.33 & 16.80$\pm$3.00 & 27$^{\dagger}$ \\ 
        EPIC 249893012 b & 1.95$\pm$0.09 & 8.75$\pm$1.09 & 6.50$\pm$1.21 & 22.57$\pm$3.50 & 24$^{*,\ddagger}$ \\ 
        EPIC 220674823 b & 1.60$\pm$0.10 & 7.72$\pm$0.80 & 10.39$\pm$2.23 & 29.58$\pm$4.80 &24$^{\dagger,\ddagger}$ \\ 
        L 98-59 c & 1.39$\pm$0.10 & 2.22$\pm$0.26 & 4.61$\pm$1.09 & 11.35$\pm$2.05 & 22$^{\dagger}$ \\
        GJ 3929 b & 1.09 $\pm$0.04 & 1.75 $\pm$0.44 & 7.45 $\pm$2.04 & 14.45 $\pm$3.78 & 22$^*$ \\ 
        TOI-431 b & 1.28$\pm$0.04 & 3.07$\pm$0.35 & 8.07$\pm$1.19 & 18.38$\pm$2.39 & 22$^{*,\ddagger}$ \\ 
        Kepler-78 b & 1.12 $\pm$0.11 & 1.97 $\pm$0.53 & 7.70 $\pm$3.11 & 15.36 $\pm$5.18 & 21$^{*,\ddagger}$ \\ 
        L 168-9 b & 1.39$\pm$0.09 & 4.60$\pm$0.56 & 9.44$\pm$2.16 & 23.36$\pm$4.15 & 20$^*$ \\ 
        K2-131 b & 1.50$\pm$0.07 & 6.30$\pm$1.40 & 10.29$\pm$2.70 & 27.47$\pm$6.62 & 19$^{*,\ddagger}$ \\ 
        Kepler-99 b & 1.48$\pm$0.08 & 6.15$\pm$1.30 & 10.46$\pm$2.79 & 27.54$\pm$6.54 & 17$^*$ \\ 
        WASP-47 e & 1.81$\pm$0.03 & 6.77$\pm$0.57 & 6.31$\pm$0.60 & 20.32$\pm$1.81 & 17$^{\dagger,\ddagger}$ \\ 
        HD 136352 b & 1.66$\pm$0.04 & 4.72$\pm$0.42 & 5.65$\pm$0.67 & 16.72$\pm$1.72 & 17$^{\dagger}$ \\ 
        Kepler-80 d & 1.53$\pm$0.09 & 6.75$\pm$0.69 & 10.39$\pm$2.12 & 28.29$\pm$4.41 & 16$^*$ \\ 
        55 Cnc e & 1.88$\pm$0.03 & 7.99$\pm$0.32 & 6.68$\pm$0.41 & 22.30$\pm$1.13 & 16$^{\dagger,\ddagger}$ \\
        KOI-1831 d & 1.13 $\pm$ 0.05 & 2.23 $\pm$ 0.67 & 8.52 $\pm$ 2.80 & 17.13 $\pm$ 5.37 & 16$^*$\\
        CoRoT-7 b & 1.68 $\pm$0.11 & 4.08 $\pm$1.02 & 4.73 $\pm$1.51 & 14.15 $\pm$4.01 & 15$^{\dagger,\ddagger}$ \\
        Kepler-60 b & 1.71$\pm$0.13 & 4.19$\pm$0.56 & 4.62$\pm$1.22 & 14.06$\pm$2.85 & 12$^{\dagger}$ \\ 
        HD 260655 c & 1.53$\pm$0.05 & 3.09$\pm$0.48 & 4.73$\pm$0.87 & 12.90$\pm$2.18 & 11$^{\dagger}$ \\ 
        Kepler-114 c & 1.60$\pm$0.18 & 2.80$\pm$0.60 & 3.77$\pm$1.51 & 10.73$\pm$3.33 & 11$^{\dagger}$ \\ 
        HD 23472 b & 2.00$\pm$0.11 & 8.32$\pm$0.78 & 5.73$\pm$1.09 & 20.41$\pm$2.95 & 10$^{\dagger}$ \\ 
        KOI-1599.01 & 1.90$\pm$0.30 & 4.60$\pm$0.30 & 3.70$\pm$1.77 & 12.50$\pm$4.03 & 9$^{\dagger}$ \\ 
        K2-111 b & 1.82$\pm$0.11 & 5.29$\pm$0.76 & 4.84$\pm$1.12 & 15.67$\pm$2.94 & 8$^{\dagger,\ddagger}$ \\ 
        GJ 367 b & 0.72$\pm$0.05 & 0.55$\pm$0.08 & 8.13$\pm$2.17 & 10.39$\pm$2.16 & 7$^{*,\ddagger}$ \\ 
        HD 137496 b & 1.31$\pm$0.06 & 4.04$\pm$0.55 & 9.90$\pm$1.92 & 23.10$\pm$3.79 & 7$^*$ \\ 
        Kepler-105 c & 1.31$\pm$0.07 & 4.60$\pm$0.92 & 11.28$\pm$2.89 & 26.30$\pm$5.96 & 6$^*$ \\ 
        Kepler-406 b & 1.43$\pm$0.03 & 6.35$\pm$1.40 & 11.97$\pm$2.74 & 30.46$\pm$6.84 & 6$^{*\ddagger}$ \\ 
        Kepler-107 c & 1.60$\pm$0.03 & 9.39$\pm$1.77 & 12.71$\pm$2.47 & 36.12$\pm$6.91 & 5$^{*,\ddagger}$ \\ 
        TOI-1634 b & 1.79$\pm$0.08 & 4.91$\pm$0.68 & 4.72$\pm$0.91 & 15.03$\pm$2.48 & 5$^{\dagger}$ \\ 
        TOI-1685 b & 1.70$\pm$0.07 & 3.78$\pm$0.63 & 4.24$\pm$0.88 & 12.83$\pm$2.39 & 3$^{\dagger}$ \\ 
        L 98-59 d & 1.52$\pm$0.12 & 1.94$\pm$0.28 & 3.04$\pm$0.84 & 8.23$\pm$1.75 & 2$^{\dagger}$ \\ 
        K2-3 b & 1.98$\pm$0.10 & 6.48$\pm$0.99 & 4.60$\pm$0.99 & 16.22$\pm$2.97 & 2$^{\dagger}$ \\ 
        TOI-776 b & 1.85$\pm$0.13 & 4.00$\pm$0.90 & 3.48$\pm$1.07 & 11.47$\pm$3.04 & 2$^{\dagger}$ \\ 
        Kepler-60 c & 1.90$\pm$0.15 & 3.85$\pm$0.81 & 3.09$\pm$0.98 & 10.46$\pm$2.75 & 1$^{\dagger}$ \\ 
        TOI-561 b & 1.42$\pm$0.07 & 1.59$\pm$0.36 & 3.04$\pm$0.81 & 7.70$\pm$1.88 & 1$^{\dagger,\ddagger}$ \\ 
        Kepler-1972 c & 0.87 $\pm$0.05 & 2.11 $\pm$0.59 & 17.78 $\pm$5.88 & 27.47 $\pm$8.33 & 1$^*$ \\ 
        Kepler-60 d & 1.99$\pm$0.16 & 4.16$\pm$0.84 & 2.91$\pm$0.92 & 10.31$\pm$2.66 & 1$^{\dagger}$ \\
        Kepler-1972 b & 0.80 $\pm$ 0.04 & 2.02 $\pm$ 0.59 & 21.58 $\pm$ 7.14 & 30.81 $\pm$ 9.55 & 0$^*$ \\ 
        HD 23472 c & 1.87$\pm$0.12 & 3.41$\pm$0.88 & 2.87$\pm$0.93 & 9.57$\pm$2.76 & 0$^{\dagger}$ \\ 
        K2-229 b & 1.00$\pm$0.02 & 2.49$\pm$0.42 & 13.72$\pm$2.46 & 24.43$\pm$4.23 & 0$^{*,\ddagger}$ \\ 
        \enddata
\tablecomments{$^{*}$Likely Fe-enriched super-Mercury $^{\dagger}$Potential volatile-rich mini-Neptune/water world (or other unexplained process to lower planet density) $^{\ddagger}$Planet has equlibrium temperature greater than zero pressure melting curve of dry peridotite \citep[$\sim$1300 K; ][]{Katz2003}.  } 
\end{deluxetable}

\subsection{Confirming planet classification using host-star abundances }
\label{sec:final}
The NRPZ is useful for broad classification of planets, particularly for demographic studies. 
Our calculated boundaries of the NRPZ (e.g., $M_{\rm MN-WW} < M  < M_{\rm SM}$) assume that host-star abundances in general represent a good baseline composition of the rock-iron portions of a planet.
Better classification of a planet---as a rocky exoplanet, super-Mercury, or mini-Neptune/water world---relies on additional constraints on its composition, typically the bulk composition of the host-star itself \citep[e.g.,][]{Dorn15, Unter16, Hink17, Bonomo19, Schulze21}. 
We explore below how to test whether our classifications based on broad stellar abundances compare when the abundances for the specific host star are available.

To do this, we use the ExoLens software package \citep{Schulze21} contained within ExoPlex to quantify the probability, $P(H_0)$, that the composition inferred from an individual rocky exoplanet's mass and radius matches its host star's composition. 
ExoLens samples within the measured mass and radius, and their respective uncertainties, to produce the distribution of possible CMFs that reproduce these measurements, ${\rm CMF}_\rho$, as determined using ExoPlex. 
ExoLens is able to utilize skewed distributions of mass and/or radius when determining ${\rm CMF}_\rho$.
Likewise, ExoLens computes the distribution of planetary CMFs predicted using the host star's abundances and their uncertainties, ${\rm CMF}_{\star}$ (Equation~\ref{eq:cmf}). 
All models assume a simple two-layer core and silicate planet with all Fe present within the core and no core light elements. 
To calculate $P(H_0)$, we quantify the amount of overlap between $CMF_{\rho}$ and $CMF_{\star}$, normalized by the null hypothesis \citep[Eq. 4 of][]{Schulze21}. 
$P(H_0)$ is therefore defined as: 
\begin{equation}
    P(H_0) = P(\text{CMF}_\rho = \text{CMF}_{\mathrm{\star}}) = \frac{\int^{1}_{0} \phi(\text{CMF}_\rho, \sigma_{\text{CMF}_\rho})\phi(\text{CMF}_{\mathrm{\star}}, \sigma_{\text{CMF}_{\mathrm{\star}}})d\text{CMF}}{\int^{1}_{0} \phi(0.5, \sigma_{\text{CMF}_\rho})\phi(0.5, \sigma_{\text{CMF}_{\mathrm{\star}}})d\text{CMF}}
    \label{equ:probability}
\end{equation}
where $\phi(\mathrm{CMF_\rho}, \sigma_{\mathrm{CMF_\rho}})$ and $\phi(\mathrm{CMF_{\star}}, \sigma_{\mathrm{CMF_{\star}}})$ are the probability distributions of $\mathrm{CMF_\rho}$ and $\mathrm{CMF_{\star}}$, respectively. 
While \citet{Schulze21} assumes the probability distributions for both $\mathrm{CMF}_\rho$ and $\mathrm{CMF_{\star}}$ are Gaussian, this assumption is not required by equation \ref{equ:probability}, and we have updated ExoLens such that we no longer make this assumption in this work.
We define the null hypothesis as both distributions having the same mean values. 
This method implicitly accounts for CMF values outside of 0 to 1, as we do not renormalize ${\rm CMF}_\rho$. 
Rather, the integral of ${\rm CMF}_\rho$ from 0 to 1 gives the pecentage of M-R pairs that can be explained by the 2-layer iron core + silicate mantle model. 
Where ${\rm CMF}_\rho < 0$, those M-R combinations require a water or volatile layer. 
Where ${\rm CMF}_\rho > 1$, the planet is unphysically dense. 
Mathematically, however, since the CMF inferred from the host-star's abundance data is bound between zero to one as required by our null hypothesis, the overlap outside of these ranges is always zero and we need only integrate from 0 to 1.
$P(H_0)$ therefore represents the degree to which we are unable to distinguish between whether a planet's measured composition is indeed that of its host star. 
In this case, as $P(H_0)$ decreases, we are increasingly confident that the planet's measured composition is statistically distinguishable from its host star's. 

\begin{deluxetable}{lccccccccccc}

\tablecolumns{12}
\tablewidth{0pt}
\tabletypesize{\scriptsize}
\tablecaption{Sample of well-characterized exoplanets with available host-star abundances. Host star elemental ratios are expressed as molar ratios derived using the solar abundances of \citet{Lodd09}. All masses and radii taken from the NASA Exoplanet Archive (\citet[][]{Akeson13}, \dataset[10.26133/NEA1]{https://doi.org/10.26133/NEA1}) and stellar abundance data are from the Hypatia Catalog. The classifications in the last column are illustrated in Figure~\ref{fig:lens}, where SM = Super-Mercury, and IHS = indistinguishable from host star. CMF values correspond to the median values and 1$\sigma$ confidence intervals.} \label{tab:plans}
\tablehead{
         \colhead{ } & \colhead{Mass} &\colhead{Radius}  & \colhead{Host} & \colhead{Host} & \colhead{Host} & \colhead{Host} & \colhead{Host} &\colhead{}  &\colhead{}  &\colhead{P(H$_{\rm{0}}$)}  &\colhead{1$\sigma$} \\
        \colhead{Planet} & \colhead{ (M$_{\oplus}$)} & \colhead{ (R$_{\oplus}$)}& \colhead{[Fe/H]} & \colhead{[Mg/H]} & \colhead{[Si/H]}& \colhead{Fe/Mg}& \colhead{Si/Mg}& \colhead{CMF$_{\star}$} & \colhead{CMF$_{\rho}$} & \colhead{(\%)}& \colhead{Class}
}

\startdata
Kepler-406 b & 6.35$\pm$1.40& 1.43$\pm$0.03 & 0.21$\pm$0.04 & 0.20$\pm$0.07 & 0.19$\pm$0.05 & 0.83$\pm$0.25 &0.93$\pm$0.26 & 0.33$\pm$0.07 & 0.76$^{+0.13}_{-0.19}$ & 8 & SM\\
Kepler-105 c & 4.60$^{+0.92}_{-0.85}$& 1.31$\pm$0.07 & -0.12$\pm$0.04 & -0.09$\pm$0.07 & -0.08$\pm$0.05 &  0.76$\pm$0.22& 0.98$\pm$0.27& 0.30$\pm$0.07 & 0.78$^{+0.18}_{-0.21}$ & 10 & SM \\
Kepler-99 b & 6.15$\pm$1.3& 1.48$\pm$0.08 & 0.26$\pm$0.04 & 0.25$\pm$0.07 & 0.21$\pm$0.10 & 0.83$\pm$0.25 &0.87$\pm$0.24 & 0.33$\pm$0.07 & 0.65$^{+0.19}_{-0.27}$ & 39 & IHS \\
Kepler-102 d & 2.5$\pm$1.40$^\dagger$& 1.18$\pm$0.04 & 0.12$\pm$0.04 & 0.08$\pm$0.07 & 0.10$\pm$0.05 & 0.88$\pm$0.26 &0.99$\pm$0.28 & 0.33$\pm$0.08 & 0.61$^{+0.31}_{>0.61}$ & 54 & IHS \\
Kepler-78 b & 1.77$^{+0.24}_{-0.25}$& 1.23$^{+0.018}_{-0.019}$ & 0.00$\pm$0.04 & -0.07$\pm$0.07 & -0.03$\pm$0.05 & 0.95$\pm$0.28 &1.05$\pm$0.29 & 0.34$\pm$0.08 & 0.14$^{+0.15}_{>0.14}$ & 60 & IHS \\
Kepler-36 b & 3.83$^{+0.11}_{-0.10}$& 1.50$^{+0.061}_{-0.049}$ & -0.18$\pm$0.04 & -0.18$\pm$0.07 & -0.19$\pm$0.05 &  0.81$\pm$0.24 &0.92$\pm$0.26 & 0.32$\pm$0.07 & 0.19$^{+0.13}_{-0.17}$ & 73 & IHS \\
Kepler-93 b & 4.54$\pm$0.85 & 1.57$\pm$0.11 & -0.18$\pm$0.04 & -0.13$\pm$0.07 & -0.19$\pm$0.05 & 0.72$\pm$0.21 & 0.83$\pm$0.23 & 0.33$\pm$0.07 & 0.16$^{+0.34}_{>0.16}$ & 93 & IHS \\
\enddata  
\tablecomments{$^{\dagger}$Mass value taken from \citet{Brink22}.}
\end{deluxetable} 

\begin{figure}
    \centering
    \includegraphics[width=\linewidth]{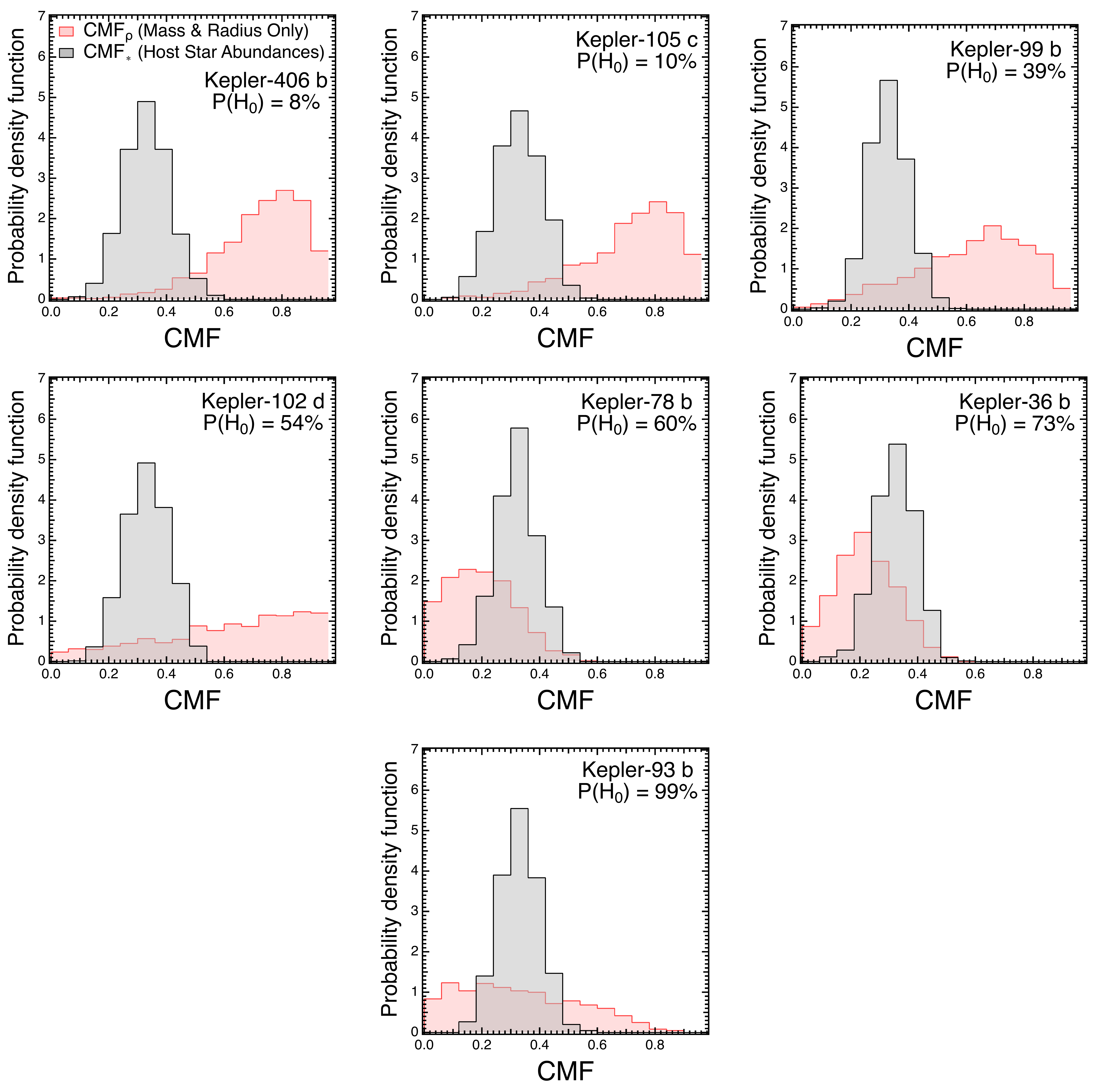}
    \caption{Probability density functions as a function of core mass fraction for CMF$_\rho$ (red) and CMF$_\star$ (gray). All CMF calculations assumes all Fe remains in the core. Average and $1\sigma$ values for both CMF distributions are listen in Table \ref{tab:plans}}.
    \label{fig:probs}
\end{figure}

We apply our methodology to a sample of seven planets orbiting FGK-type stars not included in \citet[][see Table~\ref{tab:plans}]{Schulze21}. 
These planets have average masses between 1.77 and $6.35 \, M_\oplus$, and average radii between 1.18 and $1.57 \, R_\oplus$. 
In addition, all seven host stars have stellar abundances available from the Hypatia Catalog, from \citet{Brew16, Brew18}. 
Table~\ref{tab:plans} provides the abundances [Fe/H], [Mg/H] and [Si/H] in dex units, where the uncertainties are given 
in the Hypatia Catalog
as either the \textit{spread} or the representative average error typically associated with that element, whichever is larger \citep{Hink14}. 
The stellar abundances in Table~\ref{tab:plans} (Columns 4-6) were converted to Fe/Mg and Si/Mg molar ratios (Columns 7 and 8), using the solar normalization of  \citet{Lodd09}.

\begin{figure}
    \centering
    \includegraphics[width=\linewidth]{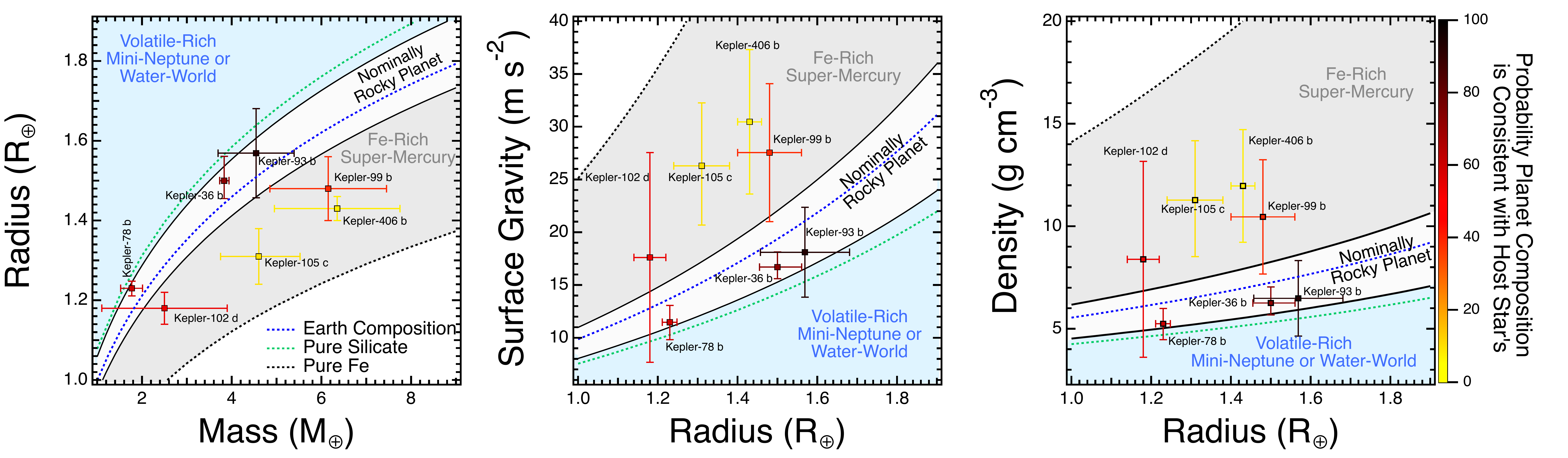}
    \caption{Radius as a function of planet mass (left), and surface gravity (center) and bulk density (right) as functions of planet radius, for the 7 planets in Table~\ref{tab:plans}. Colors represent the probability that ${\rm CMF}_{\rho}$ inferred from a planet's measured mass and radius is indistinguishable from ${\rm CMF}_{\star}$ predicted from its host star's composition. The likely super-Mercury, nominally rocky and likely mini-Neptune/ water world zones are included for reference. }
    \label{fig:lens}
\end{figure}

Comparing the predicted CMFs (Figure \ref{fig:probs}) from the host star's composition we find that two planets (Kepler-406 b and Kepler-105 c) have compositions inferred from mass and radius alone that deviate from their host stars' compositions at the $< 1\sigma$ level ($P(H_{0}) \leq 32\%$, Figure~\ref{fig:lens}). 
Each planet has inferred distributions of ${\rm CMF}_{\rho}$ (from mass and radius measurements alone) that exceed the distributions of ${\rm CMF}_{\star}$ predicted from their host star's abundances. 
This means these planets are denser than expected, confirming that they are indeed super-Mercuries as predicted from their location relative to the NRPZ. The remaining four planets (Kepler-99 b, Kepler-102 d, Kepler-78 b, Kepler-36 b and Kepler-93 b) all have $P(H_{0}) \geq 32\%$, meaning that their inferred compositions from mass and radius alone are not statistically distinguishable from that of their host star's at the $> 1\sigma$ level. 
Kepler-78 b, Kepler-36 b and Kepler-93 b are less dense than an Earth-like rocky planet composition for their mass and/or radius, and very near the lower limits in surface gravity and bulk density of the NRPZ.
While this may indicate that significant atmospheres are present, we are unable to statistically distinguish whether they are of a different composition than their host star, given all the uncertainties in mass, radius and stellar composition. 
Kepler-99 b and Kepler-102 d have average densities high enough to be considered a super-Mercury (Figure~\ref{fig:lens}). 
Their large uncertainties in mass, however, leads to them having significantly large uncertainties in their ${\rm CMF}_\rho$ distributions. 
With ${\rm CMF}_\rho$ being mostly unconstrained, we are unable to distinguish between its composition from mass and radius alone and that of its host-star at the $> 1\sigma$ level. 
We note that the inclusion of light elements within the core or mantle FeO in our models for those planets in the NRPZ are likely to only increase their respective $P(H_{0})$. 
At fixed planet mass and Fe/Mg, inclusion of light elements in the core or mantle FeO predict a lower planet density than the Fe-free mantle + LE-free core case (Figures \ref{fig:feo_per}, \ref{fig:sifeo_diff}, \ref{fig:O_core}), meaning that at a given density, the inferred $CMF_\rho$ will be larger.
We would, therefore, be less likely to differentiate whether the inferred compositions of Kepler-78 b, Kepler-36 b and Kepler-93 b from mass and radius alone are in fact distinguishable from that of their host-star's.
Examining each effect directly on these planets is the subject of future work. 

For most part, the inability to distinguish between ${\rm CMF}_{\rho}$ and ${\rm CMF}_{\star}$ stems from large uncertainties in mass, radius and/or host-star composition. 
Some planetary systems with extremely high or low density can be distinguished now. 
For example, Kepler-102 d has an uncertainty in density $\approx 137\%$, but its bulk density $12.6 \pm 6.0 \, {\rm g} \, {\rm cm}^{-3}$ puts it well into the super-Mercury zone, giving it $P(H_0) = 12\%$ and a high probability it is distinct from its host star's composition. 
On the other hand, Kepler-93 b has a similar uncertainty, $\sigma_{{\rm CMF}\rho}/{\rm CMF}_{\rho} \approx 140\%$, but with a bulk density $6.5 \pm 1.4 \, {\rm g} \, \rm cm^{-3}$, $P(H_{0})$ = 99\%, meaning it could very well be a rocky exoplanet consistent with its host star's abundances.
As uncertainties in mass and radius decrease, our ability to distinguish between a planet's inferred composition and its host star's composition will improve.

These results show that the super Mercury, volatile-rich mini-Neptune/water world, and nominally rocky planet zones in mass, radius, gravity and density allow broad characterization of small exoplanets; but accurate classification and  verification of this scheme demand measurements of the host stars' compositions.
Only then can we tell whether high-density planets have undergone mantle stripping, or simply orbit very Fe-rich stars; and whether low-density planets have volatile layers, or simply orbit Fe-poor stars.

\section{Conclusion}
Here we present the publicly available, open-source ExoPlex mass-radius-composition calculator for small exoplanets with radii up to $\sim 2 \, R_\oplus$. 
Unlike other mass-radius models, ExoPlex calculates the unique mantle mineralogy for a given bulk composition, and varying mantle/core oxidation states across a wide range of potential exoplanet composition. 
Using ExoPlex, we calculated the resulting core mass fractions (CMFs), radii, surface gravities and bulk densities for $10^6$ model planets with masses up to $\sim 15 \, M_\oplus$, with variable mantle FeO, core light element budgets and surface water contents. 
We adopted the distribution of FGK-type stellar abundances from the Hypatia Catalog as an estimate of the potential distribution of rocky exoplanet bulk compositions.
Commensurate with previous mass-radius studies, we find considerable degeneracy when inferring a planet's CMF, bulk Si/Mg ratio, oxidation state, core light element content and surface volatile budget from mass and radius alone. 
The planet's bulk Fe/Mg content, however, is potentially resolvable without additional constraints on the planet's composition. 
One such potential constraint is the composition of the planet's host star, yet few exoplanet hosts have abundance data available. 

For planets around host stars with incomplete abundance information, the range of stellar abundances can be used to define boundaries in mass-radius space between different classes of small exoplanets: super-Mercuries, rocky and volatile-rich planets (e.g., mini-Neptunes/water-worlds) . 
We modeled the distribution of rocky exoplanet radii across a wide range of interior geochemistries, and the full range of FGK-type stellar abundances from the Hypatia Catalog. 
We were able to define the boundaries in planet radius versus mass, or surface gravity or bulk density versus radius, of nominally rocky exoplanets, defining the NRPZ.
Planets definitely denser than this we define as Fe-enriched super-Mercuries.
Planets definitely less dense than this we define as volatile-rich and potentially mini-Neptunes or water worlds.
Applying the NRPZ to a sample of 85 exoplanets with both mass and radius uncertainties $\leq$30\%, we estimate nine lie within the NRPZ at greater than 70\% confidence. 
Conversely, $\sim20\%$ and $\sim$30\% of this sample of well resolved exoplanets can be classified as either Fe-enriched super-Mercuries or volatile-enriched, respectively, at $\geq$70\% confidence.

As more exoplanets are discovered, these nominally super-Mercury, mini-Neptune/water world and rocky planet zones can provide demographic studies of small exoplanets \citep[e.g., ][]{Weis14,Fulton17} with a 
simple metric for classifying individual exoplanets even without available host-star abundances. 
But we emphasize that because of the inherent degeneracies present in mass-radius models, host-star abundances are critically needed to confirm this broad compositional classification \citep[e.g., ][]{Schulze21}, yet few are available for even the best-resolved small exoplanets.
While the Hypatia Catalog does include abundance data for cooler M-dwarf stars, there have only been $\sim$800 non-Fe element abundances measured in $\sim$300 M-dwarf stars, or 1-2 elements per star. Therefore, it is not currently possible to say whether M-dwarfs abundances track FGK-type stars, especially in terms of the primary planet-building elements Mg, Si, Fe. More abundance data are therefore needed to robustly say whether the NRPZ for M-dwarf stars varies considerably from that of the FGK-type stellar sample.

We tested the likelihood that host star composition is an acceptable proxy for their planet's composition for a sample of seven planets 1.2--$1.6 \, R_\oplus$ in radius, with available host-star abundances. 
Of these seven planets, we are able to confirm that three of the densest planets (Kepler-102 d, Kepler-105 c and Kepler-406 b) have compositions and implied CMFs that are significantly different than those predicted from their measured host star's composition, at the $1\sigma$ level after including the uncertainties in all measurements; this makes them likely super-Mercuries. 
The remaining four planets (Kepler-99 b, Kepler-78-b, Kepler-36 b and Kepler-93 b) each have measured masses and radii that are statistically indistinguishable from their host stars'. 
We are thus not able to confidently conclude that the nominal super-Mercury Kepler-99 b has been enriched in iron.
Likewise, despite Kepler-78 b, Kepler-36 b and Kepler-93 b each having densities lower than a similarly sized planet with an Earth-like composition, we cannot confidently distinguish whether they are volatile-enriched or simply orbit stars likely to produce low-CMF planets.
As mass and radius uncertainties decrease we can more confidently test this hypothesis. 

Planet composition is inherently tied to its formation \citep[e.g., ][]{Bond06,Elkins08,Bond10,Wurm13,Thia15,Unter18,Dorn19, Fisch20}. 
It is only by considering planet composition holistically, by combining astronomical measurements with the thermodynamic constraints available from geophysical research, that we will be able to better quantify the likelihood a planet is Earth-like in its composition, mineralogy and potentially the manner in which it formed. 

\begin{acknowledgments}
CTU and NRH acknowledge NASA support from grant \#20-XRP20\_2-0125. The research shown here acknowledges use of the Hypatia Catalog Database, an online compilation of stellar abundance data as described in \citet{Hink14} that was supported by NASA's Nexus for Exoplanet System Science (NExSS) research coordination network and the Vanderbilt Initiative in Data-Intensive Astrophysics (VIDA). The results reported herein benefited from collaborations and/or information exchange within NASA's Nexus for Exoplanet System Science (NExSS) research coordination network sponsored by NASA's Science Mission Directorate (grant NNX15AD53G, PI Steve Desch).
\end{acknowledgments}
\software{\texttt{SciPy}, \citep{scipy}, \texttt{qhull} \citep{qhull},  \texttt{Perple\_X} \citep{Conn09}, \texttt{ExoLens}, \citep{Schulze21}}

\end{document}